%% file: Two_Way_Covert_Single.tex
\documentclass[12pt, draftclsnofoot,onecolumn,letterpaper]{IEEEtran}

\usepackage[utf8]{inputenc} 
\usepackage[T1]{fontenc}
\usepackage{url}              
\usepackage{cite}             

\usepackage[cmex10]{amsmath}  
\usepackage{mleftright}       
\mleftright                   

\usepackage{graphicx}         
\usepackage{booktabs}         
\usepackage{algorithm,algpseudocode,tabularx,diagbox}
\usepackage{subcaption}
\usepackage{todonotes}
\usepackage{etoolbox}
\usepackage{amsfonts}
\usepackage{amsmath}
\usepackage{bbm}
\usepackage{acronym}
\usepackage{amssymb}
\usepackage{comment}
\usepackage{placeins}
\usepackage{paralist}
\usepackage{xcolor}
\usepackage{hyperref}
\definecolor{navyblue}{RGB}{0, 0, 128}
\hypersetup{
  colorlinks=true,
  linkcolor=navyblue,
  citecolor=navyblue,
  filecolor=navyblue,
  urlcolor=navyblue
}

\usepackage{pgf,pgfplots}
\pgfplotsset{compat=newest}
\usepgfplotslibrary{groupplots}
\usepgfplotslibrary{dateplot}

\allowdisplaybreaks
\newtheorem{definition}{Definition}[section]
\newtheorem{theorem}[definition]{Theorem}
\newtheorem{lemma}[definition]{Lemma}
\newtheorem{proposition}[definition]{Proposition}
\newtheorem{remark}[definition]{Remark}
\newcommand{\tuna}[1]{\textcolor{black}{#1}}

\input{Acronyms.tex}

\begin{document}
\include{CommandsAndMacros.tex}

\title{Covert Communication over \\ Physically-Degraded Alarm Two-Way Channels}

\author{%
  \IEEEauthorblockN{Tuna Erdo\u{g}an, Tyler Kann, Aria Nosratinia, Matthieu Bloch}
  \thanks{This work was supported by the National Science Foundation award 1955401.}
}

\maketitle
\begin{abstract}
  We study covert communications over binary-input discrete memoryless alarm two-way channels, in which two users interact through a two-way channel and attempt to hide the presence of their communication from an eavesdropping receiver. The alarm two-way channel is one in which simultaneous transmissions by both users trigger an alarm at the eavesdropper, which captures the challenges and opportunities of cooperation beyond interference management. In particular, by characterizing the covert capacity region of two-way channels when using public time sharing, we show how cooperation strictly improves achievable covert communication throughputs. While our analysis falls short of characterizing the two-way covert capacity region for all two-way channels, we provide general achievable and converse bounds that illuminate the cooperation mechanisms that benefit covertness and are tight for a physically-degraded alarm two-way channels. Because of the unique nature of covert communications, our analysis also shows that the coordination required to avoid triggering alarms comes asymptotically ``for free.'' The key technical challenge that we address is how to appropriately design auxiliary random variables in a multi-user covert communication setting subject to the square root law.
\end{abstract}

\section{Introduction}
\label{sec:intro}

The increasing concerns over privacy in modern communication networks have highlighted the importance of not only protecting the content of transmitted signals but also hiding the presence of signals themselves. In particular, information-theoretic principles underlying steganography~\cite{Fridrich2009} have been adapted and extended to study covert communications, also known as communications with low probability of detection~\cite{Bash2013}. Covert communications are unique in that they happen in a zero-rate regime, specifically one in which the \emph{square-root law}~\cite{Bash2013} prevents the transmission of more than $\calO(\sqrt{n})$ bits over $n$ channel uses. An appropriate notion of covert capacity has been defined and characterized for point-to-point \acp{DMC} and Gaussian channels~\cite{Wang2016b,Bloch2015b} under various covertness metrics~\cite{Tahmasbi2017}, as well as classical-quantum channels~\cite{Wang2016c,Sheikholeslami2016,Bullock2025Fundamental} and \tuna{noisy} bosonic channels~\cite{Bullock2020,Gagatsos2020Covert,Wang2022Towards,Wang2024}.

The ideas underlying covert communications have also been extended to multi-user models. In the context of multiple-access~\cite{Arumugam2018a,Cho2022Covert,Bounhar2023Mixing,bounhar2024}, interference~\cite{Cho2021Treating}, and broadcast channels~\cite{Arumugam2018b,Kibloff2019,Tan2019}, the zero-rate regime of covert communications sometimes leads to surprising simplifications of the capacity regions. Intuitively, this happens because users cannot create significant interference. For state-dependent channels, it is sometimes possible to achieve positive rates of covert communications by appropriately exploiting state knowledge to align signals with channel noise~\cite{Lee2018a,ZivariFard2020} or merely using the state as jamming mechanism to confuse the eavesdropper \ac{wrt} its detection baseline~\cite{Sobers2017}. \tuna{In such cases, covert communications become similar to stealth communications~\cite{Hou_Kramer2014}.} The problems of secure and covert communication~\cite{Hou2014,Bloch2015b}, as well as covert secret-key generation~\cite{Tahmasbi2017c,Tahmasbi2018b,Tahmasbi2019a,Lin2020Stealthy} have also been studied, highlighting how to adapt information-theoretic secrecy mechanisms to the covert setting.

We are here interested in studying the problem of covert communication over two-way channels~\cite{shannon1961two}. Two-way channels are notoriously difficult to analyze because the interactions enabled by the two-way nature of the communication channel allow one to deploy a myriad of coding strategies, thereby making it challenging to identify the optimal ones. For certain classes of channels with symmetry properties, however, the capacity region is known~\cite{Weng2019Capacity}. In the context of secret communications, two-way channels have also proved to be a valuable model to capture the combination of several information-theoretic secrecy coding mechanisms, from cooperative jamming to wiretap coding and secret key generation~\cite{Tekin2008General,Tekin2008erratum,Pierrot2011a,ElGamal2013,He2013Role}. The benefits of cooperation for channel resolvability, a critical coding mechanism to enable covertness~\cite{Bloch2015b}, have also been highlighted in~\cite{Helhal2018a} for cribbing multiple-access channels. Our objective is to develop insight into how these mechanisms may specifically enhance covert communications, as well. We focus primarily on alarm two-way channels, precisely defined in Section~\ref{sec:modelandresults}, for they exacerbate the need to precisely coordinate users to avoid detection, and are motivated by situations in which introducing too much energy into a channel triggers behaviors that facilitate detection. For instance, avalanche photodiodes in optical systems display such behavior. The key contributions of the present work are
\begin{inparaenum}[i)]
\item the characterization of the covert capacity region of two-way channels under public time-sharing;
\item a coding scheme for covert communication over two-way channels, leveraging coordination for two-way communication;
\item a specific analysis of alarm two-way channels, showing how coordination may asymptotically come for free because of unique aspects of covert communications;
\item a converse region, matching the achievability for physicall-degraded alarm two-way channels.
\end{inparaenum}

The covertness requirement introduces challenges beyond those already existing in multi-user communications. Specifically, identifying auxiliary random variables is not enough, as one should also identify how the distributions of these variables change with the blocklength. The advances made in previous works on multi-user covert communications~\cite{Arumugam2018a,Cho2022Covert,Bounhar2023Mixing,Cho2021Treating,Arumugam2018b,Kibloff2019,Tan2019} were achieved in part because the specific channel models did not require the use of auxiliary random variables in the covert regime. This is not the case in the present work, and our analysis suggests an arguably unexpected and intriguing phenomenon by which the scaling of probability mass with the blocklength need not be homogeneous across random variables. 

The remainder of the paper is organized as follows. In Section~\ref{sec:notation}, we introduce the notation used throughout the paper. In Section~\ref{sec:modelandresults}, we formally introduce the problem of covert communication over discrete memoryless two-way channels. In Section~\ref{sec:prel-covert-proc}, \tuna{we develop auxiliary results regarding the scaling of information-theoretic quantities that provide insight into the type of coordination mechanisms that enable covertness}. In Section~\ref{sec:mainresults}, we provide \tuna{our main results in the form of} the covert capacity region \tuna{of physically-degraded alarm two-way channels}. In Section~\ref{sec:proofs-main-results}, we provide detailed proofs of the achievability results, including a general converse result for covert communications over discrete memoryless two-way channels. Many technical details are relegated to the appendices to streamline the presentation.

\section{Notation}
\label{sec:notation}
Random variables are denoted by uppercase letters, e.g., $X$, and their realizations are denoted by lowercase letters, e.g., $x$. Calligraphic letters represent sets. Superscripts denote the length of a sequence of symbols, while subscripts denote the indices of a symbol in the sequence, e.g., $x^n$ denotes a length $n$ sequence and $x_i$ is its $i^{\text{th}}$ index. $P_X$ and $P_{XY}$ denote probability distributions on $\calX$ and $\calX\times\calY$, respectively. For $\lambda\in[0,1]$, we define $\Bar{\lambda}\eqdef1-\lambda$.

For two distributions $P$ and $Q$ defined on the same set $\calX$, $P$ is absolutely continuous with respect to $Q$, denoted $P\tuna{\ll} Q$, if and only if for all $\calA\subseteq\calX$ $Q(\calA)=0\implies P(\calA)=0$. The relative entropy is $\D{P}{Q}\eqdef\sum_{x\in\calX}P(x)\log\frac{P(x)}{Q(x)}$, the total variation distance is $\V{P}{Q}\eqdef\frac{1}{2}\sum_x\abs{P(x)-Q(x)}$ and the $\chi_2$ distance is $\chi_2\left(P\Vert Q\right) \eqdef \sum_{x\in\calX}\frac{\left(P(x)-Q(x)\right)^2}{Q(x)}$. Throughout the paper $\log$ denotes the natural logarithm. For an i.i.d. vector whose components are distributed according to distribution $P_X$, the product distribution is denoted as $P_X^{\otimes n}(x^n)=\prod_{i=1}^nP_X(x_i)$. $\indic{\cdot}$ is the indicator function.

\section{Two-Way Covert Communication Model}
\label{sec:modelandresults}

\begin{figure}[ht]
  \centering
  \includegraphics[width=0.7\linewidth]{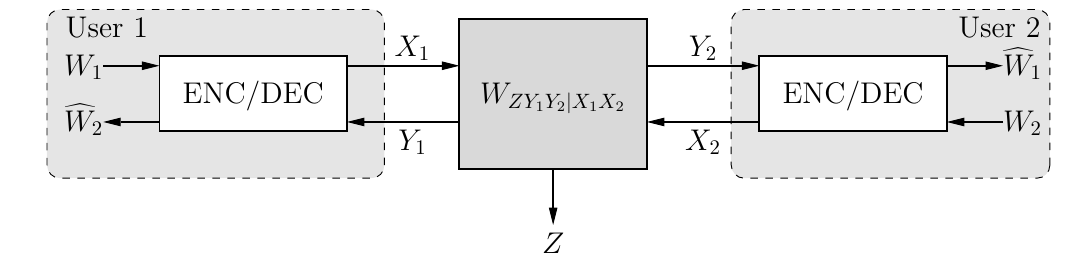}
  \caption{Two-way covert communication model.}
  \label{fig:model}
\end{figure}

We consider the communication model illustrated in Fig.~\ref{fig:model}, in which two legitimate users, User 1 and User 2, attempt to reliably communicate over a discrete memoryless two-way channel while escaping detection from an eavesdropper. The binary\footnote{The choice of binary input alphabets is not fundamental in our analysis but greatly simplifies analysis and notation.} channel input alphabet controlled by User $i\in\set{1,2}$ is $\calX_i\eqdef \set{0,1}$ while the channel output alphabet is $\calY_i$. The channel output observed by the eavesdropper is $\calZ$. All alphabets are finite, and the kernel $W_{Y_1Y_2Z|X_1X_2}$ characterizing the channel is assumed known to all parties. We assume that the symbol $0$ is an \emph{innocent} symbol that represents the absence of communication, in which case the eavesdropper expects to observe the distribution $Q_{00}\eqdef W_{Z|X_1=0,X_2=0}$. It will be convenient to also define for all $(i,j)\in\set{0,1}^2$ the distributions obtained by marginalizing the kernel:
\begin{align}
  &P^{(1)}_{ij} \eqdef W_{Y_1|X_1=i,X_2=j},\, P^{(2)}_{ij} \eqdef W_{Y_2|X_1=i,X_2=j},\, Q_{ij} \eqdef W_{Z|X_1=i,X_2=j}.
\end{align}

\begin{definition}
  Alarm two-way channels are two-way channels for which there exists an alert symbol $*\in\calZ$ such that $Q_{11}(\ast)>0$ and $Q_{ij}(\ast)=0\quad \forall (i,j)\neq(1,1)$.
\end{definition}
\begin{definition}
  \label{def:physicall-degraded}
  Physically-degraded alarm two-way channels are alarm two-way channels for which $\D{\smash{P^{(2)}_{10}}}{\smash{P^{(2)}_{00}}}>\D{Q_{10}}{Q_{00}}$ and $\D{\smash{P^{(1)}_{01}}}{\smash{P^{(1)}_{00}}}>\D{Q_{01}}{Q_{00}}$.
\end{definition}
\begin{remark}
  Although the concept of physical degradation is well established for broadcast channels~\cite{ElGamal_Kim_2011}, there is no natural extension to two-way channels, \tuna{and we acknowledge that our terminology is not standard}.
\end{remark}

In alarm two-way channels, the distribution $Q_{11}$ is not absolutely continuous with respect to $Q_{00}$, hence forbidding the two users to simultaneously use the $1$ symbol. We also assume throughout that $Q_{00}$ cannot be written as a convex combination of $\{Q_{01},Q_{10}\}$ and  that $Q_{ij},(i,j)\neq(1,1)$ is absolutely continuous \ac{wrt} $Q_{00}$. Without these assumptions, either the square root law can be circumvented or covert communication becomes impossible~\cite{Wang2016b,Bloch2015b}.

User $i\in\set{1,2}$ attempts to convey a uniformly distributed message $W_i\in\intseq{1}{M_i}$ through $n$ channel uses, possibly employing the interactions allowed by the two-way nature of the channel. At each time $t\in\intseq{1}{n}$, the transmitted symbols of User 1 and User 2 are obtained according to encoding functions
\begin{align*}
  f_{1,t}:\intseq{1}{M_1}\times\calY_1^{t-1}\to\calX_1 \text{ and }  f_{2,t}:\intseq{1}{M_2}\times\calY_2^{t-1}\to\calX_2,
\end{align*}
respectively. Encoding functions are assumed to be known to all parties, are deterministic, and do not rely on shared secret keys. The distribution induced at the eavesdropper's channel output by the use of the above encoding is denoted $\widehat{Q}^n$. After $n$ channel uses, User 1 and User 2 form estimates of the transmitted messages according to decoding functions
\begin{align*}
  g_1:\intseq{1}{M_1}\times \calY_1^n\to\intseq{1}{M_2}\text{ and }g_2:\intseq{1}{M_2}\times \calY_2^n\to\intseq{1}{M_1},
\end{align*}
respectively.

The performance of the coding scheme is assessed in terms of reliability, measured with the average probability of error
\begin{align}
  P_e^{(n)}\eqdef \P{g_1(W_1,Y_1^n)\neq W_2\text{ or }g_2(W_2,Y_2^n)\neq W_1},
\end{align}
and in terms of covertness, measured with the relative entropy $\D{\smash{\widehat{Q}^n}}{Q_{00}^{\otimes n}}$.
We refer the reader to~\cite{Bloch2015b,Tahmasbi2017} for a detailed discussion justifying the use of relative entropy as a covertness metric. Our objective is to characterize the \emph{covert capacity region} of a discrete memroyless two-way communication channel, defined next.
\begin{definition}
  \label{def:covertcapacityregion}
  For any $\delta>0$, the covert capacity region of a discrete memoryless two-way channel is the set of throughputs $(r_1,r_2)$ for which there exist sequences of coding schemes as defined earlier operating with increasingly larger blocklengths such that 
  \begin{align}
    \forall i\in\set{1,2} \quad\liminf_{n\to\infty}\frac{\log M_i}{\sqrt{n\delta}}\geq r_i,
      \label{eq:throughputdef}
  \end{align}
  and $\lim_{n\to\infty} P_e^{(n)} = 0$, $\limsup_{n\to\infty}\D{\smash{\widehat{Q}^n}}{Q_{00}^{\otimes n}} \leq \delta$.
\end{definition}
A few comments are in order to justify the non-standard definition of capacity in Definition~\ref{def:covertcapacityregion}. The number of bits transmitted is normalized by the square root of $n\delta$ in~(\ref{eq:throughputdef}) as opposed to the usual blocklength $n$. The $\sqrt{n}$ scaling reflects the existence of a square root law while $\delta$ reflects the dependence of $\log M_i$ on the covertness metric. These scalings are justified a posteriori by the results of Section~\ref{sec:mainresults} when we show that, in the limit of large blocklength, the ratios in~(\ref{eq:throughputdef}) are indeed independent of $n$ and $\delta$. The assumptions made earlier regarding the kernel $W_{ZY_1Y_2|X_1X_2}$ are crucial, as the square root law could be violated should the assumptions not be satisfied, see; e.g.,~\cite{Wang2016b,Bloch2015b}. To avoid creating confusion with the usual definition of a rate, we refer to quantities of the form $\frac{\log M}{\sqrt{n\delta}}$ as throughput instead of rate.

\section{Preliminaries: Covert Process}
\label{sec:prel-covert-proc}

Covert communication as per Definition~\ref{def:covertcapacityregion} mandates that the coding scheme achieve two objectives: non-innocent input symbols should be sparsely used and coding should introduce no noticeable structure from the eavesdropper's perspective. Following the approach in~\cite{Bloch2015b}, these two objectives are separately handled by introducing a \textit{covert process}, defined next as an \ac{iid} stochastic process $Q_Z^{\otimes n}$  indistinguishable from $Q_{00}^{\otimes n}$ in the limit of large blocklength. Conceptually, the covert process allows us to identify the fraction of non-innocent symbol that can be used and defers the problem of hiding the coding structure to the random coding argument by making connections with channel resolvability~\cite{Bloch2015b}.

In prior work on multiple access~\cite{Arumugam2018a} and interference channels~\cite{Cho2021Treating}, the covert process has typically been defined with independent random variables $X_{1}$ and $X_{2}$.\footnote{\cite{Bounhar2023Mixing,bounhar2024} introduce a public time sharing parameter that is distinct from the blocklength dependent auxiliary random variable we shall consider.} The two-way nature of the channel allows us here to make $X_{1}$ and $X_{2}$ dependent, which we capture with joint distributions of the form
\begin{align}
  P_{UX_1X_2}(u,x_1,x_2) \eqdef P_U(u)P_{X_1|U}(x_1|u)P_{X_2|U}(x_2|u)\label{eq:input_dist}.
\end{align}
In other words, we ask that an auxiliary random  variable $U$ make $X_{1}$ and $X_{2}$ conditionally independent. Note that the choice $U=(X_1,X_2)$ ensures that such an auxiliary random variable always exists, and the rationale for introducing $U$ shall become clear in Section~\ref{sec:proofs-main-results}. Suffice here to say that a specific choice of $U$ represents a specific form of coordination information passed between the two users.

Intuitively, covert communication requires that the marginal joint distribution $P_{X_1X_2}$ be low-weight, in the sense that $1-P_{X_1X_2}(0,0)=\calO\left(\frac{1}{\sqrt{n}}\right)$. This constraint, however, does not dictate how to distribute the low-weight between $P_U$ and the conditional distributions $P_{X_1|U}$ and $P_{X_2|U}$. The choice cannot be completely arbitrary, as one still needs to ensure that concentration of measure holds in the random coding arguments underlying the coding theorems. We clarify this point with two examples that illustrate this challenge unique to covert communications. 

\subsection{\tuna{Time-Sharing}}

Consider the distribution $P_{UX_1X_2}$, for $U\in\{1,2\}$, defined as
\begin{equation}
    \begin{array}{lll}
        P_U(1)=q, & P_{X_1|U}(1|1)~=p_1n^{-\frac{1}{2}}, & P_{X_2|U}(1|1)~=0,\\
        P_{U}(2)=1-q, & P_{X_1|U}(1|2)~=0, & P_{X_2|U}(1|2)~=p_2n^{-\frac{1}{2}}.
    \end{array}\label{eq:TS_dist}
\end{equation}
with $q,p_1,p_2\in [0,1]$, which corresponds to a strict time sharing between the two users. $U$ indicates when each user is allowed to transmit, leaving to each user the ability to determine how to maintain covertness. 

The corresponding eavesdropper output distribution is then
\begin{align}
    Q_Z(z) &\eqdef \sum_{x_1,x_2}P_U(u)P_{X_1|U}(x_1|u)P_{X_2|U}(x_2|u)Q_{x_1x_2}(z)\\
                            &= Q_{00}(z)+n^{-\frac{1}{2}}(qp_1+(1-q)p_2)\left(\frac{qp_1Q_{10}(z)+(1-q)p_2Q_{01}(z)}{qp_1+(1-q)p_2}-Q_{00}(z)\right).\label{eq:qz_ts1}
\end{align}
The scalings of associated information-theoretic quantities for this scheme are identified in the following lemma.
\begin{lemma}
    \label{lm:D_localTS1}
    With the choice of $P_{UX_1X_2}$ in~\eqref{eq:TS_dist}, 
    \begin{align}
      &\D{Q_Z}{Q_{00}} = n^{-1}\frac{\left(qp_1+(1-q)p_2\right)^2}{2}\chi_2\left(\frac{qp_1Q_{10}+(1-q)p_2Q_{01}}{qp_1+(1-q)p_2}\bigg\Vert Q_{00}\right)+\calO\left(n^{-\frac{3}{2}}\right),\\
      &\I{X_1}{Y_2|X_2,U} = qp_1n^{-\frac{1}{2}}\D{P^{(2)}_{10}}{P^{(2)}_{00}}+\calO\left(n^{-1}\right)\label{eq:I12},\\
      &\I{X_2}{Y_1|X_1,U} = (1-q)p_2n^{-\frac{1}{2}}\D{P^{(1)}_{01}}{P^{(1)}_{00}}+\calO\left(n^{-1}\right),\\
      &\I{X_1,X_2}{Z} = n^{-\frac{1}{2}}\left(
        qp_1\D{Q_{10}}{Q_{00}}+(1-q)p_2\D{Q_{01}}{Q_{00}}\right)+\calO\left(n^{-1}\right),\\
      &\I{X_1,U}{Z} = qp_1n^{-\frac{1}{2}}\D{Q_{10}}{Q_{00}}+\calO\left(n^{-1}\right),\\
      &\I{X_2,U}{Z} = (1-q)p_2n^{-\frac{1}{2}}\D{Q_{01}}{Q_{00}}+\calO\left(n^{-1}\right),\label{eq:I2Z}\\
      &\I{U}{Z} = \calO\left(n^{-1}\right).\label{eq:scaling_iuz_ts1}
    \end{align}
  \end{lemma}
  \begin{IEEEproof}
    See  Appendix~\ref{appendix:D_localTS1}.
  \end{IEEEproof}
  The scalings of the mutual information quantities \eqref{eq:I12}-\eqref{eq:I2Z} are on the order of $\calO\left(\smash{n^{-\frac{1}{2}}}\right)$, which seems consistent with the throughput definition in~\eqref{eq:throughputdef}. Unfortunately, one still needs to handle the random coding arguments carefully and the scaling of $\I{U}{Z}$ on the order of $\calO\left(\smash{n^{-1}}\right)$ in~(\ref{eq:scaling_iuz_ts1}) is problematic as it seems to preclude the use of concentration of measure arguments as in standard information-theoretic random coding arguments. \tuna{Specificially, as shown in Appendix~\ref{appendix:TS1},} the \tuna{$\calO\left(n^{-1}\right)$} scaling \tuna{vanishes so} fast \tuna{that it results in trivial concentration bounds}. Fortunately, this problematic scaling is circumvented using another coordination scheme.

\subsection{\tuna{Sparse Time-Sharing}}
\tuna{Consider the distribution $P_{UX_1X_2}$, for $U\in\{0,1,2\}$,} defined as
\begin{equation}
    \begin{array}{lll}
        P_U(0)=1-(q_1+q_2)n^{-\frac{1}{4}}, & P_{X_1|U}(1|0)~=0, & P_{X_2|U}(1|0)~=0,\\ P_U(1)=q_1n^{-\frac{1}{4}}, & P_{X_1|U}(1|1)~=p_1n^{-\frac{1}{4}}, & P_{X_2|U}(1|1)~=0,\\
        P_{U}(2)=q_2n^{-\frac{1}{4}}, & P_{X_1|U}(1|2)~=0, & P_{X_2|U}(1|2)~=p_2n^{-\frac{1}{4}}.
    \end{array}\label{eq:STS_dist}
\end{equation}
with $q_1,q_2,p_1,p_2\in [0,1]$, which also corresponds to a strict time sharing between the two users but distributes the low weight differently across $P_UP_{X_1|U}P_{X_2|U}$. $U$ can be thought of as an on-off keying signal that indicates when $X_1$ and $X_2$ can deviate from the innocent symbol. The eavesdropper output distribution is now
\begin{align}
    Q_Z(z) &\eqdef \sum_{x_1,x_2}P_U(u)P_{X_1|U}(x_1|u)P_{X_2|U}(x_2|u)Q_{x_1x_2}(z)\\
                            &= Q_{00}(z)+n^{-\frac{1}{2}}(q_1p_1+q_2p_2)\left(\frac{q_1p_1Q_{10}(z)+q_2p_2Q_{01}(z)}{q_1p_1+q_2p_2}-Q_{00}(z)\right).\label{eq:qz_ts2}
\end{align}
While~(\ref{eq:qz_ts1}) and~(\ref{eq:qz_ts2}) both result in mixtures of $Q_{00}$, $Q_{01}$, and $Q_{10}$ with similar weights, the crucial difference between the two is revealed when considering the associated information-theoretic quantities, as shown in Lemma~\ref{lm:D_localTS2} below.
\begin{lemma}
  \label{lm:D_localTS2}
  With the choice of $P_{UX_1X_2}$ in~\eqref{eq:STS_dist}, 
  \begin{align}
    &\D{Q_Z}{Q_{00}} = n^{-1}\frac{\left(q_1p_1+q_2p_2\right)^2}{2}\chi_2\left(\frac{q_1p_1Q_{10}+q_2p_2Q_{01}}{q_1p_1+q_2p_2}\bigg\Vert Q_{00}\right)+\calO\left(n^{-\frac{3}{2}}\right),\\
    &\I{X_1}{Y_2|X_2,U} = p_1q_1n^{-\frac{1}{2}}\D{P^{(2)}_{10}}{P^{(2)}_{00}}+\calO\left(n^{-\frac{3}{4}}\right),\\
    &\I{X_2}{Y_1|X_1,U} = p_2q_2n^{-\frac{1}{2}}\D{P^{(1)}_{01}}{P^{(1)}_{00}}+\calO\left(n^{-\frac{3}{4}}\right),\\
    &\I{X_1,X_2}{Z} = n^{-\frac{1}{2}}\left(
      q_1p_1\D{Q_{10}}{Q_{00}}+q_2p_2\D{Q_{01}}{Q_{00}}\right)+\calO\left(n^{-1}\right),\\
    &\I{X_1,U}{Z} = q_1p_1n^{-\frac{1}{2}}\D{Q_{10}}{Q_{00}}+\calO\left(n^{-\frac{3}{4}}\right),\\
    &\I{X_2,U}{Z} = q_2p_2n^{-\frac{1}{2}}\D{Q_{01}}{Q_{00}}+\calO\left(n^{-\frac{3}{4}}\right),\\
    &\I{U}{Z} = \calO\left(n^{-\frac{3}{4}}\right).\label{eq:scaling)iuz_ts2}
  \end{align}
\end{lemma}
\begin{IEEEproof}
  See  Appendix~\ref{appendix:D_localTS2}.
\end{IEEEproof}
The scaling of $\I{U}{Z}$ in~(\ref{eq:scaling)iuz_ts2}) is now $\calO(n^{-3/4})$, which will prove crucial to develop random coding arguments in Section~\ref{sec:proofs-main-results}. Also note that $\I{X_1X_2}{Z}$ scales as $\calO(n^{-1/2})$, which dominates $\I{U}{Z}$ and will also lead to intriguing features unique to covert communications.

\section{Main results}
\label{sec:mainresults}

Before we develop more sophisticated covert communication schemes, we study a public time-sharing scheme in which the two legitimate users take turns transmitting based on a public schedule known to all parties. This public time-sharing scheme serves as a baseline to assess the benefits of coordination. Without loss of generality, we assume that the first $\lambda$ fraction of the blocklength $n$ is used by User 1 to transmit coded message to User 2 while User 2 stays silent; the remaining $\Bar{\lambda}$ fraction of the blocklength is used by User 2 to transmit coded messages to User 1 while User 1 stays silent.
\begin{theorem}
  \label{thm:public-time-sharing}
  For physically-degraded two-way channels, the covert capacity region under public time-sharing is
    \begin{align}
        \calC_{\textnormal{PTS}} &= \bigcup_{\substack{\lambda\in[0,1]}}\left\{\begin{array}{l}
            (r_1,r_2)\in\bbR^2_+:\\
            r_1\leq \lambda c^{\textnormal{PTS}}_1\D{P^{(2)}_{10}}{P^{(2)}_{00}}\\
            r_2\leq \Bar{\lambda}c^{\textnormal{PTS}}_2\D{P^{(1)}_{01}}{P^{(1)}_{00}}
            \end{array}\right\},
    \end{align}
where $c^{\textnormal{PTS}}_1\eqdef\sqrt{\frac{2}{\chi_2\left(Q_{10}\Vert Q_{00}\right)}}$ and $c^{\textnormal{PTS}}_2\eqdef\sqrt{\frac{2}{\chi_2\left(Q_{01}\Vert Q_{00}\right)}}$.
\end{theorem}
\begin{IEEEproof}
  See Section~\ref{sec:achievabilityproof_PTS}.
\end{IEEEproof}
The public time-sharing effectively reduces the problem to two independent point-to-point communication channels. The point-to-point converse~\cite{Bloch2015b} shows that the physically-degraded conditions $\smash{\D{\smash{P^{(1)}_{01}}}{\smash{P^{(1)}_{00}}}}>\D{Q_{01}}{Q_{00}}$ and $\smash[]{\D{\smash{P^{(2)}_{10}}}{\smash{P^{(2)}_{00}}}}>\D{Q_{10}}{Q_{00}}$ are required to achieve non-zero throughputs for $r_1$ and $r_2$, respectively, under public time-sharing.

\begin{theorem}
  \label{thm:capacity}
    For physically-degraded alarm two-way channels, the covert capacity region is
    \begin{align}
        \calC = \bigcup_{\lambda\in[0,1]}\left\{\begin{array}{l}
            (r_1,r_2)\in\bbR^2_+:\\
            r_1 \leq \lambda c_\lambda\D{P^{(2)}_{10}}{P^{(2)}_{00}}\\
            r_2 \leq \Bar{\lambda} c_\lambda\D{P^{(1)}_{01}}{P^{(1)}_{00}}
        \end{array}\right\},
    \end{align}
    where $c_\lambda\eqdef \sqrt{\frac{2}{\chi_2\left(\lambda Q_{10}+\Bar{\lambda}Q_{01}\Vert Q_{00}\right)}}$.
  \end{theorem}
  \begin{IEEEproof}
    See Section~\ref{sec:achievabilityproof2}.
  \end{IEEEproof}
  Unlike Theorem~\ref{thm:public-time-sharing}, the physically-degraded conditions $\D{\smash{P^{(1)}_{01}}}{\smash{P^{(1)}_{00}}}>\D{Q_{01}}{Q_{00}}$ and $\D{\smash{P^{(2)}_{10}}}{\smash{P^{(2)}_{00}}}>\D{Q_{10}}{Q_{00}}$ are not necessary in Theorem~\ref{thm:capacity}. In particular, the two-way channel enables cooperation to generate secret messages and extract keys that would allow a user with a priori no channel advantage over the eavesdropper to communicate covertly, see, e.g.,~\cite{ZivariFard2020}. Although we could derive achievable throughput regions in this case, we have kept the discussion focused on alarm two-way channels for brevity.
  
\begin{figure}[ht]
    \centering
    \includegraphics[scale=0.8]{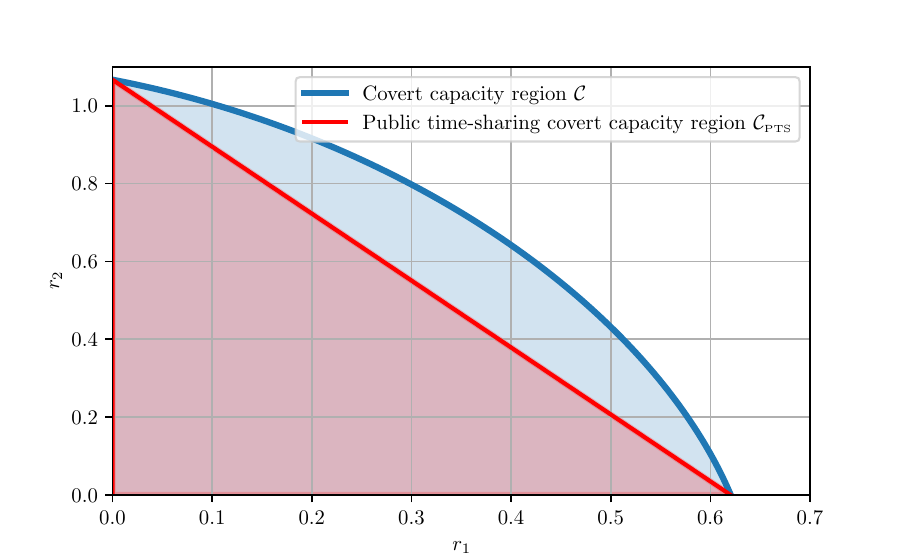}
    \caption{Comparison of the public time-sharing covert capacity region $\calC_{\textnormal{PTS}}$ and the covert capacity region $\calC$ for the example in Table~\ref{tab:P1}-Table~\ref{tab:Q}.}
    \label{fig:cap_reg}
\end{figure}

We next illustrate the results of Theorem~\ref{thm:public-time-sharing} and Theorem~\ref{thm:capacity} with a numerical example. Since $Q_{11}$, $ P^{(1)}_{10}$, $ P^{(1)}_{11}$, $ P^{(2)}_{01}$, $ P^{(2)}_{11}$ do not appear in our regions, we do not specify their values and implicitly assume that they characterize a two-way alarm channel. The values of $P^{(1)}_{X_1X_2}$, $P^{(2)}_{X_1X_2}$, and $Q_{X_1X_2}$ are given in Table~\ref{tab:P1}, Table~\ref{tab:P2}, and Table~\ref{tab:Q}, respectively, and satisfy  $\D{\smash{P^{(1)}_{01}}}{\smash{P^{(1)}_{00}}}>\D{Q_{01}}{Q_{00}}$ and $\D{\smash{P^{(2)}_{10}}}{\smash{P^{(2)}_{00}}}>\D{Q_{10}}{Q_{00}}$. The associated regions $\calC_{\textnormal{PTS}}$ and $\calC$ are shown in Figure~\ref{fig:cap_reg}, clearly illustrating the benefits of coordination for covert communications over two-way alarm channels.

\begin{table}[ht]
\centering
\caption{$P^{(1)}_{X_1X_2}$ for all $(X_1,X_2)\in\{(0,0),(0,1)\}$ and $Y_1\in\{0,1,2\}$.}
\begin{tabular}{|c|c|c|c|}
    \hline
     \backslashbox[8em]{$(X_1,X_2)$}{$Y_1$}
& 0 & 1 & 2 \\
     \hline
     (0,0) & 0.15 & 0.25 & 0.6 \\
     \hline
     (0,1) & 0.45 & 0.3 & 0.25\\
     \hline
\end{tabular}
\label{tab:P1}

\end{table}

\begin{table}[ht]
\centering
\caption{$P^{(2)}_{X_1X_2}$ for all $(X_1,X_2)\in\{(0,0),(1,0)\}$ and $Y_2\in\{0,1,2\}$.}
\begin{tabular}{|c|c|c|c|}
    \hline
     \backslashbox[8em]{$(X_1,X_2)$}{$Y_2$}
& 0 & 1 & 2 \\
     \hline
     (0,0) & 0.4 & 0.35 & 0.25 \\
     \hline
     (1,0) & 0.25 & 0.2 & 0.55\\
     \hline
\end{tabular}
\label{tab:P2}

\end{table}

\begin{table}[ht]
\centering
\caption{$Q_{X_1X_2}$ for all $(X_1,X_2)\in\{(0,0),(0,1),(1,0)\}$ and $Z\in\{0,1,2,3\}$.}
\begin{tabular}{|c|c|c|c|c|}
    \hline
     \backslashbox[8em]{$(X_1,X_2)$}{$Z$}
& 0 & 1 & 2 & 3 \\
     \hline
     (0,0) & 0.2 & 0.3 & 0.3 & 0.2 \\
     \hline
     (0,1) & 0.35 & 0.2 & 0.2 & 0.25 \\
     \hline
     (1,0) & 0.2 & 0.1 & 0.4 & 0.3\\
     \hline
\end{tabular}
\label{tab:Q}

\end{table}

\section{Proofs of Main Results}
\label{sec:proofs-main-results}
\subsection{Achievability and Converse Proof of Theorem~\ref{thm:public-time-sharing}}\label{sec:achievabilityproof_PTS}
Under public time-sharing, denote $x_1^{\lambda n}(w_1)$ (resp. $x_2^{\overline{\lambda}n}(w_2)$) the codeword encoding User 1's message $w_1$ (resp. User 2's message $w_2$). The distribution induced at the eavesdropper's channel output becomes
\begin{align}
    \widehat{Q}^n(z^n) &= \widehat{Q}^{\lambda n}_1(z^{\lambda n})\times\widehat{Q}^{\overline{\lambda}n}_2(z^{\lambda n+1:n}),\label{eq:pts_structure}
\end{align}
where
\begin{align}
    \widehat{Q}_1^{\lambda n}(z^{\lambda n}) &= \frac{1}{M_1}\sum_{w_1=1}^{M_1}W_{Z|X_1X_2}^{\otimes n}(z^{\lambda n}|x_1^{\lambda n}(w_1),0^{\lambda n}),\\
    \widehat{Q}_2^{\overline{\lambda} n}(z^{\overline{\lambda} n}) &= \frac{1}{M_2}\sum_{w_2=1}^{M_2}W_{Z|X_1X_2}^{\otimes n}(z^{\overline{\lambda} n}|0^{\overline{\lambda} n},x_2^{\overline{\lambda} n}(w_2)).
\end{align}
Now, considering the structure of the eavesdropper output distribution in~(\ref{eq:pts_structure}), our covertness metric becomes
\begin{align}
    &\D{\widehat{Q}^n}{Q_{00}^{\otimes n}} = \D{\widehat{Q}_1^{\lambda n}}{Q_{00}^{\otimes \lambda n}}+\D{\widehat{Q}_2^{\overline{\lambda} n}}{Q_{00}^{\otimes \overline{\lambda}n}}.
    \label{eq:red_ptp}
\end{align}
We introduce $\delta_1\eqdef\limsup_{n\to\infty}\D{\smash{\widehat{Q}^{\lambda n}_1}}{\smash{Q_{00}^{\otimes \lambda n}}}$ to parametrize the covertness budget allocated to User 1 during the first $\lambda n$ channel uses. Because of~\eqref{eq:red_ptp}, the two-way covert communication under public time-sharing reduces to two independent point-to-point covert communication problems. From~\cite{Bloch2015b}, in order to maximize the throughput in the point-to-point case, we need to use all of the allocated covertness budget. Hence, by incorporating this result with the definition of $\delta_1$, we shall assume that $\limsup_{n\to\infty}\D{\smash{\widehat{Q}_2^{\overline{\lambda} n}}}{\smash{Q_{00}^{\otimes \overline{\lambda} n}}}=\delta-\delta_1$ to satisfy the overall constraint
\begin{equation}
    \limsup_{n\to\infty}\D{\widehat{Q}^n}{Q_{00}^{\otimes n}}=\delta.
\end{equation}

The characterization of the covert capacity under public time sharing implicitly requires the identification of the optimal covertness budget $\delta_1$ in terms of the time-sharing parameter $\lambda$. As in~\cite{Tan2019}, we shall show that $\delta_1(\lambda)=\lambda\delta$ is optimal, though our proof is different because~\cite{Tan2019} follows a different converse argument.
\begin{proposition}
  \label{prop:achiv_pts}
  For any $\mu\in(0,1)$, $\lambda\in[0,1]$, $\delta_1\in(0,\delta)$, there exists a public time-sharing scheme such that
  \begin{align*}
        &\lim_{n\to\infty}\frac{\log M_1}{\sqrt{\lambda n\delta_1}}\geq (1-\mu)\sqrt{\frac{2}{\chi_2\left(Q_{10}\Vert Q_{00}\right)}}\D{P^{(2)}_{10}}{P^{(2)}_{00}},\\
    &\lim_{n\to\infty}\frac{\log M_2}{\sqrt{\overline{\lambda} n(\delta-\delta_1)}}\geq(1-\mu)\sqrt{\frac{2}{\chi_2\left(Q_{01}\Vert Q_{00}\right)}}\D{P^{(1)}_{01}}{P^{(1)}_{00}},\\
    &\lim_{n\to\infty} P_e^{(n)} = 0,
      \qquad \limsup_{n\to\infty} \D{\widehat{Q}^n}{Q_{00}^{\otimes n}} \leq\delta.
  \end{align*}
\end{proposition}
\begin{IEEEproof}
  By~(\ref{eq:red_ptp}), we split the covertness condition across two independent sub-blocks of length $\lambda n$ and $\overline{\lambda}n$, respectively, in which a single user is transmitting an encoded message with covertness constraint $\delta_1$, and $\delta-\delta_1$, respectively. By adapting the point-to-point covert achievability result~\cite[Theorem 2]{Bloch2015b}, there exists a public time-sharing scheme satisfying the conditions of the proposition.
\end{IEEEproof}
\begin{proposition}
  \label{prop:converse_pts}
  if $\lim_{n\to\infty} P_e^{(n)} = 0$ and $\limsup_{n\to\infty} \D{\smash{\widehat{Q}^n}}{Q_{00}^{\otimes n}} \leq\delta$, any public time-sharing scheme allocating a fraction $\lambda\in[0,1]$ of channel uses to User 1 and a covertness constraint $\delta_1\in(0,\delta)$ so that $\D{\smash{\widehat{Q}_1^{\lambda n}}}{Q_{00}^{\otimes \lambda n}}=\delta_1$ must satisfy
  \begin{align*}
        &\lim_{n\to\infty}\frac{\log M_1}{\sqrt{\lambda n\delta_1}}\leq\sqrt{\frac{2}{\chi_2\left(Q_{10}\Vert Q_{00}\right)}}\D{P^{(2)}_{10}}{P^{(2)}_{00}},\\
    &\lim_{n\to\infty}\frac{\log M_2}{\sqrt{\overline{\lambda} n(\delta-\delta_1)}}\leq\sqrt{\frac{2}{\chi_2\left(Q_{01}\Vert Q_{00}\right)}}\D{P^{(1)}_{01}}{P^{(1)}_{00}}.
  \end{align*}
\end{proposition}
\begin{IEEEproof}
  The result follows by applying the point-to-point converse result from~\cite[Theorem 3]{Bloch2015b}.
\end{IEEEproof}
Together, Proposition~\ref{prop:achiv_pts} and Proposition~\ref{prop:converse_pts} show that the public time-sharing covert capacity region is
\begin{align}
\calC_{\textnormal{PTS}} = \bigcup_{\lambda\in[0,1]}\bigcup_{\delta_1\in(0,\delta)}\left\{
  \begin{array}{l}
    (r_1,r_2)\in\bbR^2_+:\\
    r_1 \leq \sqrt{\frac{2\lambda\delta_1}{\delta\chi_2\left(Q_{10}\Vert Q_{00}\right)}}\D{P^{(2)}_{10}}{P^{(2)}_{00}}\\
    r_2\leq \sqrt{\frac{2\overline{\lambda}(\delta-\delta_1)}{\delta\chi_2\left(Q_{01}\Vert Q_{00}\right)}}\D{P^{(1)}_{01}}{P^{(1)}_{00}}
  \end{array}
  \right\}.
  \label{eq:full_charac_cpts}
\end{align}
The boundary of the region $\calC_{\textnormal{PTS}}$ can be further characterized by identifying the relationship between $\delta_1$ and $\lambda$ on the boundary as shown next.

\begin{lemma}\label{lm:opt_param}
    The parameterization of the covertness budget $\delta_1$ that corresponds to the boundary of the covert capacity region in terms of the time-sharing parameter $\lambda$ is $\delta_1(\lambda)=\lambda\delta$.
\end{lemma}
\begin{IEEEproof}
    We parameterize the boundary of the public time-sharing covert capacity region such that our free parameter for covertness is a function of our time-sharing parameter, i.e., $\delta_1=\delta_1(\lambda)$ with boundary conditions, $\delta_1(0)=0,\delta_1(1)=\delta$. These boundary conditions correspond to the maximum throughputs $r_1$ and $r_2$, respectively, that one can achieve. The area of the covert capacity region is a functional of $\delta_1$, $A(\delta_1)$  given by
\begin{align}
    A(\delta_1) &= \int r_2\mathrm{d}r_1 = \int\displaylimits_0^1 r_2(\lambda)\dot{r}_1(\lambda)\mathrm{d}\lambda
\end{align}
where the last equality follows by a change of variables. In this parameterization, we set for simplicity $r_1(\lambda,\delta_1)\eqdef\sqrt{\lambda\delta_1}$ and $r_2(\lambda_1,\delta)\eqdef\sqrt{\overline{\lambda}(\delta-\delta_1)}$, since the multiplicative terms in~(\ref{eq:full_charac_cpts}) are constants independent of the parameterization and we assume that ${\D{\smash{P^{(2)}_{10}}}{\smash{P^{(2)}_{00}}}}\geq \D{Q_{10}}{Q_{00}}$ and ${\D{\smash{P^{(1)}_{01}}}{\smash{P^{(1)}_{00}}}}\geq \D{Q_{01}}{Q_{00}}$. Hence,
\begin{align}
    \dot{r_1}(\lambda,\delta_1,\dot{\delta}_1) &= \frac{\partial r_1}{\partial \lambda}+\frac{\partial r_1}{\partial \delta_1}\dot{\delta}_1= \frac{1}{2}\left(\sqrt{\frac{\delta_1}{\lambda}}+\dot{\delta}_1\sqrt{\frac{\lambda}{\delta_1}}\right).
\end{align}

To maximize our area functional, we use the following fundamental result from calculus of variations, reduced to the one-dimensional case,
\begin{theorem}[{\cite[Theorem 1]{gelfand2012calculus}}] \label{th:EL}
    Let $\calP(a,b,x_a,x_b)$ be the set of smooth paths $q:[a,b]\to\bbR$ for which $q(a)=x_a$, $q(b)=x_b$ and let $\calL(t,q,v)$ be the Lagrangian, a smooth real-valued function. The action functional $S:\calP(a,b,x_a,x_b)\to\bbR$ is defined as
    \begin{align}
        S[q] = \int\displaylimits_a^b\calL(t,q(t),\dot{q}(t))\mathrm{d}t.
    \end{align}
    A path $q\in\calP(a,b,x_a,x_b)$ is a stationary point, extremal, of $S$ if and only if it satisfies the Euler-Lagrange equation,
    \begin{align}
        \frac{\partial\calL}{\partial q}(t,q(t),\dot{q}(t))-\frac{\mathrm{d}}{\mathrm{d}t}\frac{\partial\calL}{\partial \dot{q}}(t,q(t),\dot{q}(t))=0.\label{eq:EL}
    \end{align}
  \end{theorem}
  
Using the notation of Theorem~\ref{th:EL}, the Lagrangian of the area functional $A(\delta_1)$ is
\begin{align}
    \calL(\lambda,\delta_1,\dot{\delta}_1) = \frac{1}{2}\left(\sqrt{\frac{\delta_1}{\lambda}}+\dot{\delta}_1\sqrt{\frac{\lambda}{\delta_1}}\right)\sqrt{\overline{\lambda}(\delta-\delta_1)}.
\end{align}
The derivatives of the Lagrangian are,
\begin{align}
    \frac{\partial \calL}{\partial \delta_1}(\lambda,\delta_1,\dot{\delta}_1) &= \frac{1}{4}\left(\sqrt{\frac{\overline{\lambda}(\delta-\delta_1)}{\lambda\delta_1}}-\frac{\dot{\delta}_1}{\delta_1}\sqrt{\frac{\lambda\overline{\lambda}(\delta-\delta_1)}{\delta_1}}-\sqrt{\frac{\overline{\lambda}\delta_1}{\lambda(\delta-\delta_1)}}-\dot{\delta}_1\sqrt{\frac{\lambda\overline{\lambda}}{\delta_1(\delta-\delta_1)}}\right),\label{eq:gradL}\\
    \frac{\partial \calL}{\partial\dot{\delta}_1}(\lambda,\delta_1,\dot{\delta}_1) &= \frac{1}{2}\sqrt{\frac{\lambda\overline{\lambda}(\delta-\delta_1)}{\delta_1}},\\
    \frac{\mathrm{d}}{\mathrm{d}\lambda}\frac{\partial \calL}{\partial\dot{\delta}_1}(\lambda,\delta_1,\dot{\delta}_1) &= \frac{1}{4}\left(\sqrt{\frac{\overline{\lambda}(\delta-\delta_1)}{\lambda\delta_1}}-\sqrt{\frac{\lambda(\delta-\delta_1)}{\overline{\lambda}\delta_1}}-\frac{\dot{\delta}_1}{\delta_1}\sqrt{\frac{\lambda\overline{\lambda}(\delta-\delta_1)}{\delta_1}}-\dot{\delta}_1\sqrt{\frac{\lambda\overline{\lambda}}{\delta_1(\delta-\delta_1)}}\right).\label{eq:derL}
\end{align}
By substituting \eqref{eq:gradL} and \eqref{eq:derL} into the Euler-Lagrange equation \eqref{eq:EL}, we obtain after simplification
\begin{align}
    \sqrt{\frac{\overline{\lambda}\delta_1}{\lambda(\delta-\delta_1)}}=\sqrt{\frac{\lambda(\delta-\delta_1)}{\overline{\lambda}\delta_1}},
\end{align}
which reduces to $\delta_1=\lambda\delta$. By Theorem~\ref{th:EL} the parameterization $\delta_1(\lambda)=\lambda\delta$ is the unique extremal of the area functional $A$.
\end{IEEEproof}
Substituting the characterization provided by Lemma~\ref{lm:opt_param} in~(\ref{eq:full_charac_cpts}) yields the result of Theorem~\ref{thm:public-time-sharing}.

\subsection{Achievability Proof of Theorem~\ref{thm:capacity}}\label{sec:achievabilityproof2}
We shall prove the following achievability result.
\begin{proposition}
  \label{prop:achievability_proof2}
  For any $\mu\in(0,1)$ and  $\lambda\in[0,1]$, there exists a sequence of coding schemes such that
  \begin{align*}
    &r_1\geq \left(1-\mu\right)c_\lambda\lambda\D{P^{(2)}_{10}}{P^{(2)}_{00}},\\
    &r_1\leq \left(1+\mu\right)c_\lambda\lambda\D{Q_{10}}{Q_{00}},\\
    &r_2\geq \left(1-\mu\right)c_\lambda\overline{\lambda}\D{P^{(1)}_{01}}{P^{(1)}_{00}},\\
    &r_2\leq \left(1+\mu\right)c_\lambda\overline{\lambda}\D{Q_{01}}{Q_{00}},\\
    &\lim_{n\to\infty} P_e^{(n)} = 0,
      \qquad \limsup_{n\to\infty} \D{\widehat{Q}^n}{Q_{00}^{\otimes n}} \leq\delta,
  \end{align*}
  where $c_\lambda\eqdef \sqrt{\frac{2}{\chi_2\left(\lambda Q_{10}+\Bar{\lambda}Q_{01}\Vert Q_{00}\right)}}$.
\end{proposition}
\begin{IEEEproof}
  The general approach is inspired by~\cite{Helhal2018a}. We use a block Markov coding scheme with $B$ blocks of length $n$, where $B$ is a constant independent of $n$. In each block $b\in\intseq{1}{B}$, we analyze reliability and covertness assuming the existence of a common secret message $W_0^{(b)}$ that facilitates coordination. To justify the existence of $W_0^{(b)}$, we require each user $i\in\set{1,2}$ to split its message  $W_i^{(b)}$ into a secret part $W_i^{s,(b)}$, to be kept hidden from the eavesdropper, and a public part $W_i^{p,(b)}$. Since both users eventually get to reconstruct both $W_1^{s,(b)}$ and $W_2^{s,(b)}$, the pair $(W_1^{s,(b)},W_2^{s,(b)})$ can serve as the common secret message $W_0^{(b+1)}$ in block $b+1$. This mode of operation creates a dependence across blocks, which is illustrated by the functional dependence graph in Fig.~\ref{fig:cod_dep}. The crux of our proof is to first show how to achieve reliability, covertness, and secrecy in each block $b$ and then show how the dependence across blocks does not compromise reliability and covertness.
\begin{figure}[h]
    \centering
    \includegraphics[width=0.7\linewidth]{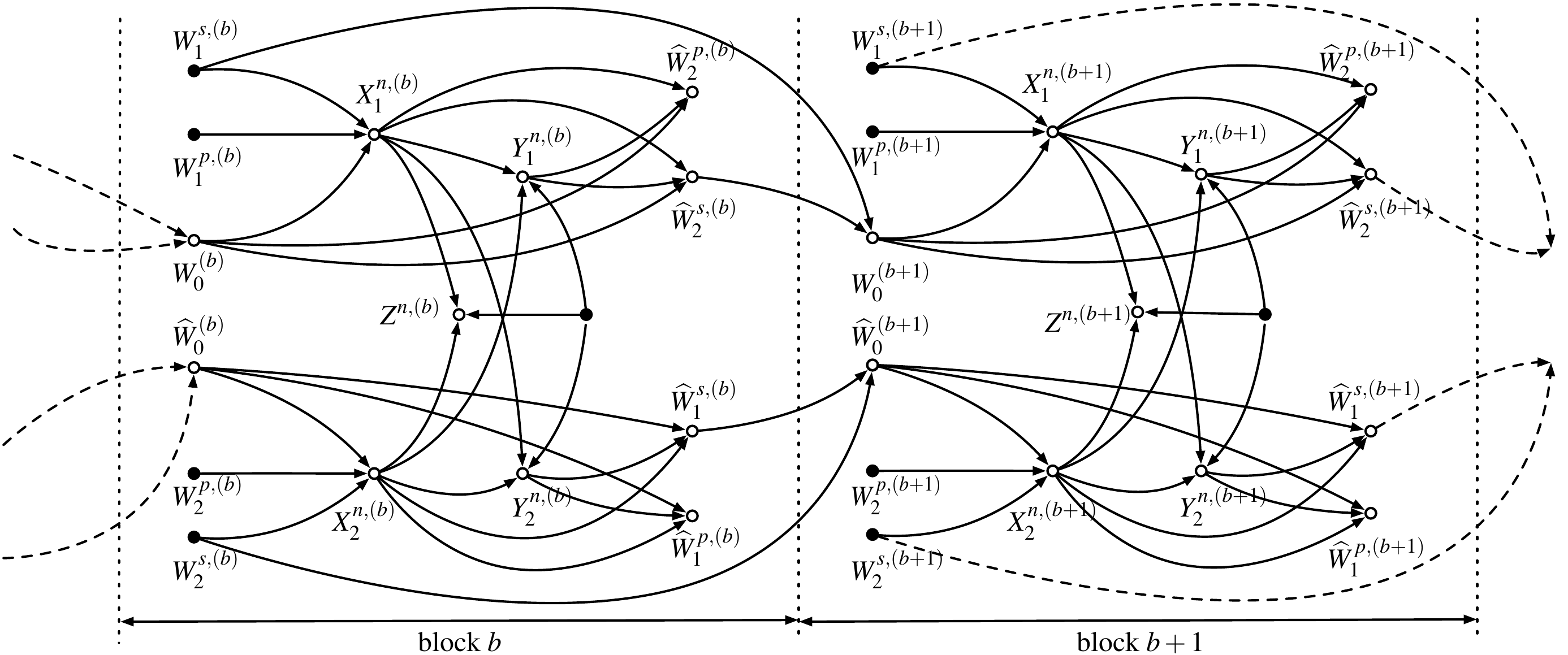}
    \caption{Functional dependence graph of the proposed coding scheme. In every block $b$, User $i\in\set{1,2}$ transmits a secret message $W_i^{s,(b)}$ and a public message $W_i^{p,(b)}$ using a common message $W_0^{(b)}$ for coordination. The codewords sent through the channel are $X^{n,(b)}_{1}$ and $X^{n,(b)}_{2}$, respectively, resulting in received signals $Y^{n,(b)}_{1}$ and $Y^{n,(b)}_{2}$ at the users and $Z^{n,(b)}$ at the eavesdropper. The secret messages are used to create the common message $W_0^{(b+1)}$ in the next block.}
    \label{fig:cod_dep}
\end{figure}

\textit{a) Random codebook generation and decoder:} Consider the distributions $P_{UX_1X_2}$ and $Q_Z$ defined in~\eqref{eq:STS_dist} and~(\ref{eq:qz_ts2}). For every block $b\in\intseq{1}{B}$:
\begin{itemize}
    \item Independently generate $M_0$ codewords according to $P_U^{\otimes n}$ and label them $u^n(w_0^{(b)})$ where $w_0^{(b)}\in\intseq{1}{M_0}$.
    \item For every $w_0^{(b)}$, independently generate $M_1^pM_1^s$ codewords according to $\prod_{i=1}^nP_{X_1|U=u^n_i(w_0^{(b)})}$; label them $x_1^n(w_1^{p,(b)},w_1^{s,(b)},w_0^{(b)})$ for $w_1^{p,(b)}\in\intseq{1}{M_1^p}$ and $w_1^{s,(b)}\in\intseq{1}{M_1^s}$. Set $w_1^{(b)}\eqdef(w_1^{p,(b)},w_1^{s,(b)})$.
    \item For every $w_0^{(b)}$, independently generate $M_2^pM_2^s$ codewords according to $\prod_{i=1}^nP_{X_2|U=u^n_i(w_0^{(b)})}$; label them $x_2^n(w_2^{p,(b)},w_2^{s,(b)},w_0^{(b)})$ for $w_2^{p,(b)}\in\intseq{1}{M_2^p}$ and $w_2^{s,(b)}\in\intseq{1}{M_2^s}$. Set $w_2^{(b)}\eqdef(w_2^{p,(b)},w_2^{s,(b)})$.
\end{itemize}
Define the codebook $\calC_n^{(b)}$ for block $b$ as
\begin{multline}
    \calC_n^{(b)} \eqdef \left\{u^n(w_0^{(b)}),x_1^n(w_1^{p,(b)},w_1^{s,(b)},w_0^{(b)}),x_2^n(w_2^{p,(b)},w_2^{s,(b)},w_0^{(b)})):\right.\\ \left.w_0^{(b)}\in\intseq{1}{M_0},w_1^{p,(b)}\in\intseq{1}{M_1^p},w_1^{s,(b)}\in\intseq{1}{M_1^s},w_2^{p,(b)}\in\intseq{1}{M_2^p},w_2^{s,(b)}\in\intseq{1}{M_2^s}\right\}.
\end{multline}
The random codebook is denoted by $C_n^{(b)}$. The transmission of codewords $x_1^n(w_1^{p,(b)},w_1^{s,(b)},w_0^{(b)})$ and $x_2^n(w_2^{p,(b)},w_2^{s,(b)},w_0^{(b)})$ in block $b$ over the discrete memoryless two-way channel results in observations $y_1^{n,(b)}$, $y_2^{n,(b)}, z^{n,(b)}$ at User 1, User 2, and the eavesdropper, respectively.

For a fixed $\mu\in(0,1)$, the decoders in every block rely on threshold decoding with the sets
\begin{align*}
  \calA_{1,\gamma_1}^n &\eqdef \left\{\begin{array}{l}\left(x_1^n,x_2^n,y_2^n,u^n\right)\in\calX_1^n\times\calX_2^n\times\calY_2^n\times\calU^n:\\
  \log\frac{W_{Y_2|X_1X_2}^{\otimes n}(y_2^n|x_1^n,x_2^n)}{P_{Y_2|X_2U}^{\otimes n}(y_2^n|x_2^n,u^n)}\geq \gamma_1\end{array}\right\},\\
  \calA_{2,\gamma_2}^n &\eqdef \left\{\begin{array}{l}\left(x_1^n,x_2^n,y_1^n,u^n\right)\in\calX_1^n\times\calX_2^n\times\calY_1^n\times\calU^n:\\
  \log\frac{W_{Y_1|X_1X_2}^{\otimes n}(y_1^n|x_1^n,x_2^n)}{P_{Y_1|X_2U}^{\otimes n}(y_1^n|x_1^n,u^n)}\geq \gamma_2\end{array}\right\},
\end{align*}
with $\gamma_1\eqdef(1-\mu)n\I{X_1}{Y_2|X_2,U}$ and $\gamma_2\eqdef(1-\mu)n\I{X_2}{Y_1|X_1,U}$. Specifically, knowing $w_0^{(b)}$ and $w_2^{(b)}$, User $2$ decodes $w_1^{(b)}$ as $w_1$ if $w_1$ is the unique message $w_1\in\intseq{1}{M_1}$ such that $(x_1^n(w_1, w_0^{(b)}), x_2^n(w_2^{(b)}, w_0^{(b)}), y_2^n, u^n(w_0^{(b)})) \in \mathcal{A}_{1,\gamma_1}^n$; else, User 2 declares a decoding error. Similarly, knowing $w_0^{(b)}$ qnd $w_1^{(b)}$, User $1$ decodes $w_2^{(b)}$ as $w_2$ if $w_2$ is the unique message $w_2\in\intseq{1}{M_2}$ such that $(x_1^n(w_1^{(b)}, w_0^{(b)}), x_2^n(w_2, w_0^{(b)}), y_1^n, u^n(w_0^{(b)})) \in \mathcal{A}_{2,\gamma_2}^n$; else, User 1 declares a decoding error.

\textit{b) Channel reliability analysis in every block:} The following holds.
\begin{lemma}
    \label{lm:reliability2}
    For any $\mu\in(0,1)$, $n$ large enough, and every block $b\in\intseq{1}{B}$, if the common message $W_0^{(b)}$ is known by legitimate users and if
    \begin{align}
        &\log M_1 < (1-\mu)q_1p_1n^{\frac{1}{2}}\D{P^{(2)}_{10}}{P^{(2)}_{00}},\\
        &\log M_2 < (1-\mu)q_2p_2n^{\frac{1}{2}}\D{P^{(1)}_{01}}{P^{(1)}_{00}},
    \end{align}
    the decoding error probability $P_e^{(b)}\eqdef \P{(\widehat{W}_1^{(b)},\widehat{W}_2^{(b)})\neq (W_1^{(b)},W_2^{(b)})}$ averaged over all random codebooks satisfies  $\E{P_e^{(b)}}\leq\exp(-\zeta n^{\frac{1}{2}})$ for an appropriate constant $\zeta>0$.
\end{lemma}
\begin{IEEEproof}
  The proof follows ideas similar to~\cite{Helhal2018a} but requires a one-shot analysis to properly handle the dependence of distributions with the blocklength. In addition, our choice of threshold decoding, which circumvents the need to characterize entropies and only relies on mutual information is crucial to obtain the result and avoid dealing with the different scaling of entropies with the blocklength. See Appendix~\ref{appendix:proof_reliability2} for details. 
\end{IEEEproof}

\textit{c) Channel resolvability analysis in every block:} The following holds. assuming a common message $W_0^{(b)}$.
\begin{lemma}
    \label{lm:resolvability2}
    For any $\mu\in(0,1)$, $n$ large enough, and every block $b\in\intseq{1}{B}$, if the common message $W_0^{(b)}$ is known by legitimate users and if
    \begin{align}
      &\log M_1^pM_2^pM_0 > (1+\mu)\left(q_1p_1n^{\frac{1}{2}}\D{Q_{10}}{Q_{00}}+q_2p_2n^{\frac{1}{2}}\D{Q_{01}}{Q_{00}}\right),\\
      &\log M_2^pM_0 > (1+\mu)q_2p_2n^{\frac{1}{2}}\D{Q_{01}}{Q_{00}},\\
      &\log M_1^pM_0 > (1+\mu)q_1p_1n^{\frac{1}{2}}\D{Q_{10}}{Q_{00}},\\
      &\log M_0 > \calO\left(n^{\frac{1}{4}}\right).
    \end{align}
    the relative entropy between $P_{Z^n|W_1^{s,(b)}W_2^{s,(b)}}$ and $Q_Z^{\otimes n}$ averaged over secret messages and all random codebooks satisfies
    \begin{align}
        \E{\E[W_1^{s,(b)}W_2^{s,(b)}]{\D{P_{Z^n|W_1^{s,(b)}W_2^{s,(b)}}}{Q_Z^{\otimes n}}}}\leq\exp\left(-\xi n^{\frac{1}{4}}\right),\label{eq:resolve_code}
    \end{align}
    for an appropriate constant $\xi>0$.
\end{lemma}
\begin{IEEEproof}
     The proof follows again ideas similar to~\cite{Helhal2018a} but requires a one-shot analysis to properly handle the dependence of distributions with the blocklength. See Appendix~\ref{appendix:proof_resolvability2} for details.
  \end{IEEEproof}
Ignoring the superscript $(b)$ for clarity, note that we control $\E[W_1^sW_2^s]{\D{P_{Z^n|W_1^sW_2^s}}{Q_Z^{\otimes n}}}$ instead of $\D{P_{Z^n}}{Q_Z^{\otimes n}}$ to ensure that $W_1^{s}$ and $W_2^{s}$ remain secret with respect to the eavesdropper observing $Z^n$. This follows by deriving ``secrecy from resolvability''~\cite{Hayashi2006,Bloch2011e} and observing that $\I{\smash{W_1^{s},W_2^{s}}}{Z^n}\leq \E[W_1^sW_2^s]{\D{P_{Z^n|W_1^sW_2^s}}{Q_Z^{\otimes n}}}$.

\textit{c) Chaining multiple blocks:} While Lemma~\ref{lm:reliability2} and Lemma~\ref{lm:resolvability2} assume far that both users have access to the common message $W_0^{(b)}$ in every block $b\in\intseq{1}{B}$, we now consider the actual operation of the protocol. After each block $b$, each user has access to their own secret message and estimate of the other user's secret message, i.e., User 1 has access to $W_1^{s,(b)}$ and $\widehat{W}_2^{s,(b)}$ and User 2 has access to $W_2^{s,(b)}$ and $\widehat{W}_1^{s,(b)}$, which they use to generate $W_0^{(b+1)}$ and $\widehat{W}_0^{(b+1)}$, respectively. This procedure enforces the additional constraint
\begin{align}
  \log M_1^sM_2^s \geq \log M_0.\label{eq:rate_chaining_constraint}
\end{align}

By the union bound, the overall error rate $P^{(nB)}_e $over all blocks can be upper bounded as
\begin{align}
  P^{(nB)}_e \leq \sum_{b=1}^B P_e^{(b)},\label{eq:union_bound}
\end{align}
Furthermore, the following lemma shows that we can control the relative entropy between the induced distribution and the covert stochastic process by the relative entropy between the induced distribution and the covert stochastic process in each block.
\begin{lemma}
  The relative entropy between the induced distribution $\widehat{Q}^{nB}$ and the covert stochastic process $Q_{Z}^{\otimes nB}$ is upper bounded by
  \begin{align}
    \D{\widehat{Q}^{nB}}{Q_{Z}^{\otimes nB}}\leq &2\sum_{b=1}^B\E[W_1^{s,(b)}W_2^{s,(b)}]{\D{P_{Z_b^n|W_1^{s,(b)},W_2^{s,(b)}}}{Q_{Z}^{\otimes n}}}\nonumber\\
    &+\sum_{b=1}^B\left(\He{\widehat{W}_1^{s,(b)}|W_1^{s,(b)}}+\He{\widehat{W}_2^{s,(b)}|W_2^{s,(b)}}\right),\label{eq:chaining_relative_entropy}
  \end{align}
  where $\widehat{W}_i^{s,(b)}$ and $W_i^{s,(b)}$ denote the estimated and the original secret message of User $i$ in block $b$.
  \label{lm:chaining}
\end{lemma}

The proof of Lemma~\ref{lm:chaining} is similar to~\cite{Helhal2018a} and is omitted for brevity. By Fano's inequality, note that
\begin{align}
\forall i\in\set{1,2}\quad  \He{\widehat{W}_i^{s,(b)}|W_i^{s,(b)}} \leq \Hb{P_e^{(b)}}+P_e^{(b)}\log M_i^s,
\end{align}
which ensures that controlling reliability and resolvability in each block shall guarantee reliability and resolvability across the entire chain of blocks. 

\textit{d) Identification of a specific code:} Using the union bound and Markov's inequality, we obtain 
\begin{align}
  \P{P_e^{(nB)}<4\E{P_e^{(nB)}}\cap\D{\widehat{Q}^{nB}}{Q_{Z}^{\otimes nB}}<4\E{\D{\widehat{Q}^{nB}}{Q_{Z}^{\otimes nB}}}}\geq\frac{1}{2}.
\end{align}
Hence, if one satisfies the constraints of Lemma~\ref{lm:reliability2}, Lemma~\ref{lm:resolvability2}, and~(\ref{eq:rate_chaining_constraint}), we conclude combining~(\ref{eq:union_bound}) and~(\ref{eq:chaining_relative_entropy}) that there must exist at least one codebook that simultaneously achieves \begin{align}
    P^{(nB)}_e &\leq\exp\left(-\zeta_1n^\frac{1}{2}\right)\label{eq:code_rel},\\
    \D{\widehat{Q}^{nB}}{Q_Z^{\otimes nB}} &\leq\exp\left(-\zeta_2n^{\frac{1}{4}}\right).\label{eq:code_res}
\end{align}  
for appropriate constants $\zeta_1,\zeta_2>0$ and an $n$ large enough.

\textit{e) Covertness:} Next, we show that we can indeed ensure covertness.
\begin{lemma}\label{lm:resolv_concentration2}
  For $n$ large enough and an appropriate constant $\zeta_3>0$, the specific code identified earlier satisfies
  \begin{align}
    \abs{\D{\widehat{Q}^{nB}}{Q_{00}^{\otimes nB}}-\D{Q_{Z}^{\otimes nB}}{Q_{00}^{\otimes nB}}} \leq \exp\left(-\zeta_3n^\frac{1}{4}\right)\label{eq:resolv_concentration}.
  \end{align}
    provided that \eqref{eq:resolve_code} is satisfied.
\end{lemma}
\begin{IEEEproof}
    See Appendix~\ref{appendix:resolv_concent_proof}.
\end{IEEEproof}
Combining Lemma~\ref{lm:D_localTS2} and Lemma~\ref{lm:resolv_concentration2}, and defining $\lambda\eqdef \frac{q_1p_1}{q_1p_1+q_2p_2}\in[0;1]$, we obtain
\begin{align}
    \D{\widehat{Q}^{nB}}{Q_{00}^{\otimes nB}} = B\frac{\left(q_1p_1+q_2p_2\right)^2}{2}\chi_2\left(\lambda Q_{10}+\Bar{\lambda}Q_{01}\Vert Q_{00}\right)+\calO\left(n^{-\frac{1}{2}}\right).\label{eq:D_res_local2}
\end{align}
Note that $\lambda$ is a function of the parameters $q_1,q_2,p_1,p_2$. Since $q_1,q_2,p_1,p_2\in[0,1]$ are free parameters, for each $\lambda$ we can find appropriate values so that 
\begin{align}
    B\frac{\left(q_1p_1+q_2p_2\right)^2}{2}\chi_2\left(\lambda Q_{10}+\Bar{\lambda}Q_{01}\Vert Q_{00}\right) &= \delta,\label{eq:cov_cons_sat}
\end{align}
and we satisfy the covertness constraint $\limsup_{n\to\infty}\D{\smash{\widehat{Q}^{nB}}}{Q_{00}^{\otimes nB}} = \delta$.

Combining \eqref{eq:code_rel},~\eqref{eq:code_res}, and~(\ref{eq:D_res_local2}), we conclude that there exists one coding scheme that simultaneously satisfies reliability and covertness constraints. In light of the constraints of Lemma~\ref{lm:reliability2}, Lemma~\ref{lm:resolvability2},~(\ref{eq:rate_chaining_constraint}), and~(\ref{eq:cov_cons_sat}), we have shown the existence of a code such that 
\begin{align}
    \lim_{n\to\infty}\frac{\log M_1}{\sqrt{nB\delta}} &\geq (1-\mu)\sqrt{\frac{2}{\chi_2\left(\lambda Q_{10}+\Bar{\lambda}Q_{01}||Q_{00}\right)}}\lambda\D{P^{(2)}_{10}}{P^{(2)}_{00}},\\
  \lim_{n\to\infty}\frac{\log M_2}{\sqrt{nB\delta}} &\geq (1-\mu)\sqrt{\frac{2}{\chi_2\left(\lambda Q_{10}+\Bar{\lambda}Q_{01}||Q_{00}\right)}}\Bar{\lambda}\D{P^{(1)}_{01}}{P^{(1)}_{00}},\\
  \lim_{n\to\infty}\frac{\log M_1^pM_0}{\sqrt{nB\delta}} &\leq (1+\mu)\sqrt{\frac{2}{\chi_2\left(\lambda Q_{10}+\Bar{\lambda}Q_{01}||Q_{00}\right)}}\lambda\D{Q_{10}}{Q_{00}},\\
  \lim_{n\to\infty}\frac{\log M_2^pM_0}{\sqrt{nB\delta}} &\leq (1+\mu)\sqrt{\frac{2}{\chi_2\left(\lambda Q_{10}+\Bar{\lambda}Q_{01}||Q_{00}\right)}}\Bar{\lambda}\D{Q_{01}}{P^{(2)}_{00}},\\
  \lim_{n\to\infty}\frac{\log M_1^pM_2^pM_0}{\sqrt{nB\delta}} &\leq (1+\mu)\sqrt{\frac{2}{\chi_2\left(\lambda Q_{10}+\Bar{\lambda}Q_{01}||Q_{00}\right)}}\left(\lambda\D{Q_{10}}{Q_{00}}+\Bar{\lambda}\D{Q_{01}}{Q_{00}}\right),\\
  \lim_{n\to\infty}\frac{\log M_0}{\sqrt{nB\delta}} &= 0.
\end{align}
Note that the constraints are now solely parameterized by $\lambda$. In addition, because $\frac{\log M_0}{\sqrt{nB\delta}}$ vanishes, its effect disappears asymptotically resulting in the region stated in Proposition~\ref{prop:achievability_proof2}.
\end{IEEEproof}

\subsection{Converse Proof of Theorem~\ref{thm:capacity}}
The converse proof of Theorem~\ref{thm:capacity} follows from a general converse for covert communication over any discrete memoryless two-way channel we describe next.
\begin{proposition}
  \label{prop:converse}
  The covert capacity region of any discrete memoryless two-way channel is included in
  \begin{align}
    \bigcup_{\substack{
    \bfrho\in[0,1]^4,\rho_{00}=0,\\ \sum_{ij}\rho_{ij}=1}}\left\{\begin{array}{l}
    (r_1,r_2)\in\bbR^2_+:\\
    r_1 \leq \tau_{\bfrho} \left(\sum_{ij}\rho_{ij}\D{P^{(2)}_{ij}}{P^{(2)}_{00}}-\left(\sum_i\rho_{i1}\right)\D{\frac{\sum_j\rho_{j1}P^{(2)}_{j1}}{\sum_k\rho_{k1}}}{P^{(2)}_{00}}\right)\\
    r_2 \leq \tau_{\bfrho} \left(\sum_{ij}\rho_{ij}\D{P^{(1)}_{ij}}{P^{(1)}_{00}}-\left(\sum_i\rho_{1i}\right)\D{\frac{\sum_j\rho_{1j}P^{(1)}_{1j}}{\sum_k\rho_{1k}}}{P^{(1)}_{00}}\right)
    \end{array}\right\},\label{eq:outerbound}
\end{align}
where $\bfrho\eqdef\set{\rho_{00},\rho_{01},\rho_{10},\rho_{11}}$ and $\tau_{\bfrho} = \sqrt{\frac{2}{\chi_2\left(\sum_{ij}\rho_{ij}Q_{ij}\middle\Vert Q_{00}\right)}}$.
\end{proposition}

\begin{remark}
Our converse result does not include lower bounds on the throughputs, which one could expect from other multi-user covert communication results~\cite{Arumugam2018a,Cho2021Treating}. While we were able to obtain single-letterized bounds of the form
\begin{align*}
    \frac{1}{n}\log M_1M_2 \geq \I{X_1X_2}{Z}-\He{Y_1Y_2|X_1X_2Z}-\epsilon,
\end{align*}
such bounds are useless in the context of covert communications because the mutual information term scales as $\calO(1/\sqrt{n})$ while the entropy scales as $\calO(1)$. Intuitively, the challenge is that $\He{Y_1Y_2|X_1X_2Z}$ captures the potential reduction in channel resolvability throughput enabled by secret coordination between the two legitimate users, but without accounting for the communication required to achieve this coordination. Unlike two-way wiretap channels, in which secrecy is an explicit constraint that can be used in the converse, secrecy is not explicitly required in the context of covert communications, suggesting that more sophisticated (and still elusive) arguments are required to develop refined converse bounds. 
\end{remark}

\begin{IEEEproof}
  The proof proceeds in two steps, first by showing how the covertness constraint induces a constraint on the average number of non-innocent symbols that can be tolerated at the channel inputs, and then showing how this constraint can be integrated into the single-letterization process. The latter part presents new challenges because of the two-way nature of the channel.

\textit{a) Relative Entropy Analysis:} Let $P_{X_{1i}X_{2i}}$ denote the joint distribution of the $i^{th}$ symbols of the codewords for User 1 and User 2 and $P_{\Tilde{X}_{1}\Tilde{X}_{2}}$ be the average joint distribution of the codewords, i.e.,
\begin{align}
    P_{X_{1i}X_{2i}}(x_1,x_2) &= \frac{1}{M_1M_2}\sum_{w_1=1}^{M_1}\sum_{w_2=1}^{M_2}\indic{x_{1i}(w_1)=x_1,x_{2i}(w_2)=x_2}\eqdef \mu^{(n),i}_{x_1x_2},\\
    P_{\Tilde{X}_1\Tilde{X}_2}(x_1,x_2) &= \frac{1}{n}\sum_{i=1}^nP_{X_{1i}X_{2i}}(x_1,x_2)\eqdef\mu^n_{x_1x_2}.
\end{align}
From the output distribution $\widehat{Q}^n$ of the eavesdropper, the marginal distribution of the $i^{th}$ symbol and time-averaged distribution are
\begin{align*}
  \widehat{Q}^n_i(z_i) &= \sum_{x_1,x_2}P_{X_{1i}X_{2i}}(x_1,x_2)Q_{x_1x_2}(z_i) = \sum_{k,\ell}\mu_{k\ell}^{(n),i}Q_{k\ell}(z_i),\\
  \overline{Q}^n(z) &= \frac{1}{n}\sum_{i=1}^n\widehat{Q}^n_i(z)
                      = \sum_{k,\ell}\mu^n_{k\ell}Q_{k\ell}(z)\quad \forall n\in\bbN^*.
\end{align*}
Since $  \limsup_{n\to\infty}\D{\smash{\widehat{Q}^n}}{Q_{00}^{\otimes n}}\leq \delta$, one can show following by~\cite{Wang2016b} that, for $n$ large enough, 
\begin{align}
  \delta &\geq \sum_{i=1}^n \D{\widehat{Q}_i^n}{Q_{00}}
         \stackrel{(a)}\geq n\D{\overline{Q}^n}{Q_{00}},\label{eq:D_convexity}
\end{align}
where $(a)$ follows from the convexity of relative entropy in the first argument. This also implies that 
\begin{align}
  \lim_{n\to\infty}\overline{Q}^n(z) &= Q_{00}(z)\quad\forall z\in\calZ.
\end{align}
From this result and by our assumptions that $Q_{00}$ cannot be written as a convex combination of $\{Q_{01},Q_{10},Q_{11}\}$ we must have
$\mu^n_{01},\mu^n_{10},\mu^n_{11} \xrightarrow[n\to\infty]{} 0$.

To obtain the scaling of the weights we use the following lemma.
\begin{lemma}\label{lm:D_conv}
  If $\lim_{n\to\infty}\mu_{ij}^n=0$ $\forall(i,j)\neq(0,0)$, the average weight of the codewords satisfies for $n$ large enough
  \begin{align}
    \D{\widehat{Q}^n}{Q_{00}^{\otimes n}} &\geq n\chi_2\left(\sum_{ij}\mu_{ij}^nQ_{ij}\middle\Vert Q_{00}\right)+\calO\left(\mu_n^3\right).\label{eq:D_LB}
  \end{align}
\end{lemma}
\begin{IEEEproof}
    See Appendix~\ref{appendix:D_conv}.
\end{IEEEproof}

\textit{b) Throughput Analysis:} For User 1 we upper bound $\log M_1$ by standard techniques,
\begin{align}
  \log M_1 &\overset{}{\leq} \I{W_1}{W_2Y_2^n}+\Hb{P_e^{(n)}}+P_e^{(n)}\log M_1\\
           &\leq \sum_{i=1}^n\I{X_{1i}}{Y_{2i}|X_{2i}}+n\epsilon_n\label{eq:M1_bound}
\end{align}
with $\epsilon_n\eqdef \frac{1}{n}(\Hb{P_e^{(n)}}+P_e^{(n)}\log M_1)$. To upper bound the first term in \eqref{eq:M1_bound} we use the following lemma.
\begin{lemma}\label{lm:conv_r1r2}
  If $\forall(k,\ell)\neq~(0,0),\quad\forall n\in\bbN^*$ $\lim_{n\to\infty}\mu_{k\ell}^{(n),i}=0$, we have
  \begin{align}
    \sum_{i=1}^n\I{X_{1i}}{Y_{2i}|X_{1i}} \leq& n\mu_n\left(\sum_{ij}\rho_{ij}^n\D{P^{(2)}_{ij}}{P^{(2)}_{00}}-\sum_i\rho_{i1}^n\D{\frac{\sum_j\rho^n_{j1}P^{(2)}_{j1}}{\sum_k\rho^n_{k1}}}{P^{(2)}_{00}}\right)\nonumber\\
    &+\calO\left(n\left(\mu_{10}^n\right)^2\right),\label{eq:I_avg_bound}
  \end{align}
  where $\forall n\in\bbN^*$ $\mu_n\eqdef 1-\mu^n_{00}$ and $\forall(i,j)$ $\rho_{ij}^n\in[0,1]^4$ we have $\rho_{00}^n=~0$, $\sum_{ij}\rho_{ij}^n=1$, $\forall(i,j)\neq~(0,0)~\mu_{ij}^n=\mu_n\rho^n_{ij}$.
\end{lemma}
\begin{IEEEproof}
See Appendix~\ref{appendix:conv_r1r2}. 
\end{IEEEproof}
Hence, by combining \eqref{eq:D_LB}, \eqref{eq:M1_bound} and \eqref{eq:I_avg_bound} we obtain
\begin{align}
  &\liminf_{n\to\infty}\frac{\log M_1}{\sqrt{n\delta}}\leq \tau_{\bfrho} \left(\sum_{ij}\rho_{ij}\D{P^{(2)}_{ij}}{P^{(2)}_{00}}-\sum_i\rho_{i1}\D{\frac{\sum_j\rho_{j1}P^{(2)}_{j1}}{\sum_k\rho_{k1}}}{P^{(2)}_{00}}\right).
\end{align}
The bound for the throughput of User 2 follows similarly.
\end{IEEEproof}

The converse proof of Theorem~\ref{thm:capacity} is the specialization of the general converse of Proposition~\ref{prop:converse} to alarm channels. For the alarm channel, covertness enforces $\rho_{11}=0$, which results in the parameterization $\rho_{10}=\lambda$, $\rho_{01}=\Bar{\lambda}$ and $\tau_{\bfrho}=c_\lambda$. With this parameterization the converse region becomes $r_1\leq c_\lambda\lambda\D{\smash{P^{(2)}_{10}}}{\smash{P^{(2)}_{00}}},r_2\leq c_\lambda\Bar{\lambda}\D{\smash{P^{(2)}_{10}}}{\smash{P^{(2)}_{00}}}$.

\bibliographystyle{IEEEtran}
\bibliography{references.bib}

\appendices
\section{Proof of Lemma~\ref{lm:D_localTS1}}\label{appendix:D_localTS1}
For the Time-Sharing joint distribution defined in~(\ref{eq:TS_dist}) and~(\ref{eq:qz_ts1}),
\begin{align}
    Q_{Z} &= Q_{00}+qp_1n^{-\frac{1}{2}}\left(Q_{10}-Q_{00}\right)+(1-q)p_2n^{-\frac{1}{2}}\left(Q_{01}-Q_{00}\right)\\
                         &= Q_{00}+n^{-\frac{1}{2}}\left(qp_1+(1-q)p_2\right)\left(\frac{qp_1Q_{10}+(1-q)p_2Q_{01}}{qp_1+(1-q)p_2}-Q_{00}\right).
\end{align}
Define $Q_{q,\vecp}\eqdef\frac{qp_1Q_{10}+(1-q)p_2Q_{01}}{qp_1+(1-q)p_2}$ where $\vecp=[p_1\,p_2]^T$,
\begin{align}
    Q_{Z} = Q_{00}+n^{-\frac{1}{2}}(qp_1+(1-q)p_2)\left(Q_{q,\vecp}-Q_{00}\right).
\end{align}
Let $\zeta(z)\eqdef Q_{q,\vecp}(z)-Q_{00}(z)$. From the definition of $\D{Q_{Z}}{Q_{00}}$, we have
\begin{align}
  \D{Q_{Z}}{Q_{00}} &= \sum_zQ_{Z}(z)\log\frac{Q_{Z}(z)}{Q_{00}(z)}\\
                    &= \sum_zQ_{00}(z)\left(1+n^{-\frac{1}{2}}(qp_1+(1-q)p_2)\frac{\zeta(z)}{Q_{00}(z)}\right)\nonumber\\
                    &\phantom{======}\times\log\left(1+n^{-\frac{1}{2}}(qp_1+(1-q)p_2)\frac{\zeta(z)}{Q_{00}(z)}\right).\label{eq:DQ}
\end{align}
Since $\log(1+x)<x-\frac{x^2}{2}+\frac{x^3}{3}$, for $x>-1$, \eqref{eq:DQ} is upper bounded by
\begin{align}
  \D{Q_{Z}}{Q_{00}} \leq &\sum_z Q_{00}(z)\left(1+n^{-\frac{1}{2}}(qp_1+(1-q)p_2)\frac{\zeta(z)}{Q_{00}(z)}\right)\nonumber\\ &\phantom{==}\times\left(n^{-\frac{1}{2}}(qp_1+(1-q)p_2)\frac{\zeta(z)}{Q_{00}(z)}-n^{-1}\frac{\left(qp_1+(1-q)p_2\right)^2}{2}\frac{\zeta(z)^2}{Q_{00}(z)^2}\right.\nonumber\\
                         &\phantom{==}\left.+n^{-\frac{3}{2}}\frac{\left(qp_1+(1-q)p_2\right)^3}{3}\frac{\zeta(z)^3}{Q_{00}(z)^3}\right)\\
                         &\overset{(a)}{=} \sum_z \left(n^{-1}\frac{\left(qp_1+(1-q)p_2\right)^2}{2}\frac{\zeta(z)^2}{Q_{00}(z)}-n^{-\frac{3}{2}}\frac{\left(qp_1+(1-q)p_2\right)^3}{6}\frac{\zeta(z)^3}{Q_{00}(z)^2}\right.\nonumber\\
                         &\phantom{==}\left.+n^{-2}\frac{\left(qp_1+(1-q)p_2\right)^4}{3}\frac{\zeta(z)^4}{Q_{00}(z)^3}\right)\\
                         &= n^{-1}\frac{\left(qp_1+(1-q)p_2\right)^2}{2}\chi_2\left(Q_{q,\vecp}||Q_{00}\right)+\calO\left(n^{-\frac{3}{2}}\right),
                           \label{eq:Dub}
\end{align}
where $(a)$ follows from the fact that $\sum_z\zeta(z)=0$. There exists $n$ large enough such that $n^{-\frac{1}{2}}\frac{\zeta(z)}{Q_{00}(z)}\geq-\frac{1}{2}$ for all $z\in\calZ$. Using the inequalities $\log(1+x)>x-\frac{x^2}{2}$, for $x>0$ and $\log(1+x)>x-\frac{x^2}{2}+\frac{2x^3}{3}$, for $x\in\left[-\frac{1}{2},0\right]$, for $n$ large enough \eqref{eq:DQ} is lower bounded by
\begin{align}
  \D{Q_{Z}}{Q_{00}} \geq& \sum_{z:\zeta(z)>0}Q_{00}(z)\left(1+n^{-\frac{1}{2}}\left(qp_1+(1-q)p_2\right)\frac{\zeta(z)}{Q_{00}(z)}\right)\nonumber\\
                        &\phantom{==}\times\left(n^{-\frac{1}{2}}\left(qp_1+(1-q)p_2\right)\frac{\zeta(z)}{Q_{00}(z)}-n^{-1}\frac{\left(qp_1+(1-q)p_2\right)^2}{2}\frac{\zeta(z)^2}{Q_{00}(z)^2}\right)\nonumber\\
                           &\phantom{==}+\sum_{z:\zeta(z)\leq0}Q_{00}(z)\left(1+n^{-\frac{1}{2}}\left(qp_1+(1-q)p_2\right)\frac{\zeta(z)}{Q_{00}(z)}\right)\nonumber\\
                        &\phantom{==}\times\left(n^{-\frac{1}{2}}\left(qp_1+(1-q)p_2\right)\frac{\zeta(z)}{Q_{00}(z)}-n^{-1}\frac{\left(qp_1+(1-q)p_2\right)^2}{2}\frac{\zeta(z)^2}{Q_{00}(z)^2}\right.\nonumber\\
                           &\phantom{==}\left.+n^{-\frac{3}{2}}\frac{2\left(qp_1+(1-q)p_2\right)^3}{3}\frac{\zeta(z)^3}{Q_{00}(z)^3}\right)\\
                           &= n^{-1}\frac{\left(qp_1+(1-q)p_2\right)^2}{2}\chi_2\left(Q_{q,\vecp}||Q_{00}\right)+\calO\left(n^{-\frac{3}{2}}\right).
                           \label{eq:Dlb}
\end{align}
Combining \eqref{eq:Dub} and \eqref{eq:Dlb} we conclude that for $n$ sufficiently large.
\begin{align}
  \D{Q_{Z}}{Q_{00}} &= n^{-1}\frac{\left(qp_1+(1-q)p_2\right)^2}{2}\chi_2\left(Q_{q,\vecp}||Q_{00}\right)+\calO\left(n^{-\frac{3}{2}}\right).
\end{align}
For $\I{X_1}{Y_2|X_2,U}$, we obtain
\begin{align}
    \I{X_1}{Y_2|X_2,U} &= \sum_{u,x_1,x_2,y_2}P_{U}(u)P_{X_1|U}(x_1|u)P_{X_2|U}(x_2|u)P^{(2)}_{x_1x_2}(y_2)\log\frac{P^{(2)}_{x_1x_2}(y_2)}{P_{Y_2|X_2U}(y_2|x_2,u)}\\
                       &= qp_1n^{-\frac{1}{2}}\D{P^{(2)}_{10}}{P^{(2)}_{00}}+\calO\left(n^{-1}\right).
\end{align}
By symmetry of $X_1$ and $X_2$,
\begin{align}
    \I{X_2}{Y_1|X_1,U} &= (1-q)p_2n^{-\frac{1}{2}}\D{P^{(1)}_{01}}{P^{(1)}_{00}}+\calO\left(n^{-1}\right).
\end{align}
For $\I{X_1,X_2}{Z}$, we obtain
\begin{align}
    \I{X_1,X_2}{Z} &= \sum_{u,x_1,x_2,z}P_U(u)P_{X_1|U}(x_1|u)P_{X_2|U}(x_2|u)Q_{x_1x_2}(z)\log\frac{Q_{x_1x_2(z)}}{Q_{00}(z)}\nonumber-\D{Q_Z}{Q_{00}}\\
                   &= n^{-\frac{1}{2}}\left(qp_1\D{Q_{10}}{Q_{00}}+(1-q)p_2\D{Q_{01}}{Q_{00}}\right)+\calO\left(n^{-1}\right).
\end{align}
For $\I{X_1,U}{Z}$, we obtain
\begin{align}
  \I{X_1,U}{Z} &= \sum_{u,x_1,x_2,z}P_{U}(u)P_{X_1|U}(x_1|u)P_{X_2|U}(x_2|u)Q_{x_1x_2}(z)\log\frac{\sum_{x_2'}P_{X_2|U}(x_2'|u)Q_{x_1x_2'}(z)}{Q_{00}}\nonumber\\
               &\phantom{===============================}-\D{Q_Z}{Q_{00}}\\
                 &= n^{-\frac{1}{2}}qp_1\D{Q_{10}}{Q_{00}}+(1-q)\D{Q_{00}+p_2n^{-\frac{1}{2}}(Q_{01}-Q_{00})}{Q_{00}}+\calO\left(n^{-1}\right)\\
                 &= n^{-\frac{1}{2}}qp_1\D{Q_{10}}{Q_{00}}+\calO\left(n^{-1}\right).
\end{align}
By symmetry of $X_1$ and $X_2$,
\begin{align}
    \I{X_2,U}{Z} &= n^{-\frac{1}{2}}(1-q)p_2\D{Q_{01}}{Q_{00}}+\calO\left(n^{-1}\right).
\end{align}
Finally, for $\I{U}{Z}$, we obtain
\begin{align}
    \I{U}{Z} &= \sum_{u,x_1,x_2,z}P_{U}(u)P_{X_1|U}(x_1|u)P_{X_2|U}(x_2|u)Q_{x_1x_2}(z)\nonumber\\
    &\phantom{==}\times\log\frac{\sum_{x_1',x_2'}P_{X_1|U}(x_1'|u)P_{X_2|U}(x_2'|u)Q_{x_1'x_2'}(z)}{Q_{Z}(z)}\\
             &= q\D{Q_{00}+p_1n^{-\frac{1}{2}}(Q_{10}-Q_{00})}{Q_{00}}\nonumber\\
             &\phantom{==}+(1-q)\D{Q_{00}+p_2n^{-\frac{1}{2}}(Q_{01}-Q_{00})}{Q_{00}}+\calO\left(n^{-1}\right)\\
             &= \calO\left(n^{-1}\right).
\end{align}

\section{Proof of Lemma~\ref{lm:D_localTS2}}\label{appendix:D_localTS2}
For the Sparse Time-Sharing scheme joint distribution defined in~(\ref{eq:STS_dist}) and~(\ref{eq:qz_ts2}), we obtain
\begin{align}
    \D{Q_Z}{Q_{00}} &= \D{Q_{00}+n^{-\frac{1}{2}}(q_1p_1+q_2p_2)\left(\frac{q_1p_1Q_{10}+q_2p_2Q_{01}}{q_1p_1+q_2p_2}-Q_{00}\right)}{Q_{00}}\\
                    &= n^{-1}\frac{(q_1p_1+q_2p_2)^2}{2}\chi_2\left(\frac{q_1p_1Q_{10}+q_2p_2Q_{01}}{q_1p_1+q_2p_2}\Vert Q_{00}\right)+\calO\left(n^{-\frac{3}{2}}\right),
\end{align}
following steps similar to Lemma~\ref{lm:D_localTS1}. For $\I{X_1}{Y_2|X_2,U}$, we obtain
\begin{align}
    \I{X_1}{Y_2|X_2,U} &= q_1n^{-\frac{1}{4}}\I{X_1}{Y_2|X_2=0,U=1}\\
                       &= q_1n^{-\frac{1}{4}}\sum_{x_1,y_2}P_{X_1|U}(x_1|1)P^{(2)}_{x_10}(y_2)\left(\log\frac{P^{(2)}_{x_10}(y_2)}{P^{(2)}_{00}(y_2)}\right.\nonumber\\
                       &\phantom{==}\left.-\log\frac{P^{(2)}_{00}(y_2)+p_1n^{-\frac{1}{4}}(P^{(2)}_{10}(y_2)-P^{(2)}_{00}(y_2))}{P^{(2)}_{00}(y_2)}\right)\\
                       &= q_1p_1n^{-\frac{1}{2}}\D{P^{(2)}_{10}}{P^{(2)}_{00}}\nonumber\\
                       &\phantom{==}-q_1p_1n^{-\frac{1}{4}}\D{P^{(2)}_{00}+p_1n^{-\frac{1}{4}}\left(P^{(2)}_{10}-P^{(2)}_{00}\right)}{P^{(2)}_{00}}\\
                       &= q_1p_1n^{-\frac{1}{2}}\D{P^{(2)}_{10}}{P^{(2)}_{00}}+\calO\left(n^{-\frac{3}{4}}\right),\label{eq:I1_local}
\end{align}
where \eqref{eq:I1_local} follows from steps similar to Lemma~\ref{lm:D_localTS1}. By symmetry of $X_1$ and $X_2$,
\begin{align}
    \I{X_2}{Y_1|X_1,U} &= q_2p_2n^{-\frac{1}{2}}\D{P^{(1)}_{01}}{P^{(1)}_{00}}+\calO\left(n^{-\frac{3}{4}}\right).
\end{align}
For $\I{X_1,X_2}{Z}$, we obtain
\begin{align}
    \I{X_1,X_2}{Z} &= \E{\D{Q_{X_1X_2}}{Q_Z}}\\
                   &= \E{\E{\D{Q_{X_1X_2}}{Q_{00}}|U}}-\D{Q_Z}{Q_{00}}\\
                   &= q_1n^{-\frac{1}{4}}\E{\D{Q_{X_10}}{Q_{00}}|U=1}+q_2n^{-\frac{1}{4}}\E{\D{Q_{0X_2}}{Q_{00}}|U=2}\nonumber\\
                   &\phantom{==}+\calO\left(n^{-1}\right)\\
                   &= q_1p_1n^{-\frac{1}{2}}\D{Q_{10}}{Q_{00}}+q_2p_2n^{-\frac{1}{2}}\D{Q_{01}}{Q_{00}}+\calO\left(n^{-1}\right).
\end{align}
For $\I{X_1,U}{Z}$, we obtain,
\begin{align}
    \I{X_1,U}{Z} &= \E{\D{\E[X_2]{Q_{X_1X_2}|U}}{Q_Z}}\\
                 &= q_1p_1n^{-\frac{1}{2}}\D{Q_{10}}{Q_{00}}+q_2n^{-\frac{1}{4}}\D{Q_{00}+p_2n^{-\frac{1}{4}}(Q_{01}-Q_{00})}{Q_{00}}\nonumber\\
                 &\phantom{==}-\D{Q_Z}{Q_{00}}\\
                 &= q_1p_1n^{-\frac{1}{2}}\D{Q_{10}}{Q_{00}}+\calO\left(n^{-\frac{3}{4}}\right).
\end{align}
By symmetry,
\begin{align}
    \I{X_2,U}{Z} &= q_2p_2n^{-\frac{1}{2}}\D{Q_{01}}{Q_{00}}+\calO\left(n^{-\frac{3}{4}}\right).
\end{align}
Finally, for $\I{U}{Z}$, we obtain
\begin{align}
    \I{U}{Z} &= \E{\D{\E{Q_{X_1X_2}|U}}{Q_{00}}}-\D{Q_Z}{Q_{00}}\\
             &= q_1n^{-\frac{1}{4}}\D{Q_{00}+p_1n^{-\frac{1}{4}}(Q_{10}-Q_{00})}{Q_{00}}\nonumber\\
             &\phantom{==}+q_2n^{-\frac{1}{4}}\D{Q_{00}+p_2n^{-\frac{1}{4}}(Q_{01}-Q_{00})}{Q_{00}}+\calO\left(n^{-1}\right)\\
             &= \calO\left(n^{-\frac{3}{4}}\right),
\end{align}
where the last equality follows from steps similar to Lemma~\ref{lm:D_localTS1}.

\section{Bernstein's Inequality}
\label{sec:bernst-ineq}
\begin{lemma}
  Let $\{I_i\}_{i=1}^n$ be independent zero-mean random variables such that $\abs{I_i}\leq c$ for a finite $c>0$ almost surely $\forall i\in\intseq{1}{n}$. Then, $\forall t>0$,
  \begin{align}
    \P{\sum_{i=1}^nI_i>t}\leq \exp\left(-\frac{\frac{1}{2}t^2}{\sum_{i=1}^n\E{I_i^2}+\frac{1}{3}ct}\right).
  \end{align}
\end{lemma}

\section{Proof of Lemma~\ref{lm:reliability2}}
\label{appendix:proof_reliability2}
In this section, we omit the block index $b\in\intseq{1}{B}$ for clarity. At the beginning of each block, User $i\in\set{1,2}$ encodes its message $w_i$ using the common message $w_0$ into a codeword $x_i^n(w_i,w_0)$ and transmits it over the discrete memoryless two-way channel. The observations of User $i\in\set{1,2}$ at the end of the block consists of the vector $y_i^n$. By the union bound, note that
\begin{align}
  \E{P_e^n} &= \P{\{\widehat{W_1}\neq W_1\}\cup\{\widehat{W_2}\neq W_2\}}\leq \P{\widehat{W_1}\neq W_1}+\P{\widehat{W_2}\neq W_2}.\label{eq:UB_error}
\end{align}
and we can analyze each user decoding error separately. We shall focus on the decoding error of $W_1$, which is determined by the following two error events:
\begin{itemize}
\item the message $w_1$ does not satisfy the threshold, i.e., $(x_1^n(w_1,w_0),x_2^n(w_2,w_0),y_2^n)\not\in\calA_{1,\gamma_1}^n$.
\item there exists another message $w_1'\neq w_1$ such that $(x_1^n(w_1',w_0),x_2^n(w_2,w_0),y_2^n)\in\calA_{1,\gamma_1}^n$.
\end{itemize}
Define the event
\begin{align}
  \calE_{w_1}^1 &\eqdef \left\{(x_1^n(w_1,w_0),x_2^n(w_2,w_0),y_2^n)\in\calA_{1,\gamma_1}^n\right\}.
\end{align}
The probability of decoding error for $W_1$ averaged over the all random codebooks then satisfies
\begin{align}
  &\P{\widehat{W}_1\neq W_1}\\
  &= \E{\frac{1}{M_1M_2M_0}\sum_{w_1,w_2,w_0}\sum_{y_2} W^{\otimes n}_{Y_2|X_1X_2}(y_2^n| x_1^n(w_1,w_0),x_2^n(w_2,w_0)) \mathbbm{1}\biggl\{\mathcal{E}_{w_1}^c \cup \bigcup_{w_1'\neq w_1}\mathcal{E}_{w_1'} \biggl\}}\label{eq:Pe1_exact}\\
                            &\overset{(a)}{\leq} \E{\frac{1}{M_1M_2M_0}\sum_{w_1,w_2,w_0}\sum_{y_2} W^{\otimes n}_{Y_2|X_1X_2}(y_2^n| x_1^n(w_1,w_0),x_2^n(w_2,w_0))\indic{\mathcal{E}^c_{w_1}}} \nonumber \\
                            &\nonumber\\
  &\phantom{===}+\E{\frac{1}{M_1M_2M_0}\sum_{w_1,w_2,w_0}\sum_{w_1'\neq w_1}\sum_{y_2} W^{\otimes n}_{Y_2|X_1X_2}(y_2^n| x_1^n(w_1,w_0),x_2^n(w_2,w_0))\indic{\mathcal{E}_{w_1'}}},\label{eq:Pe_b}
\end{align}
where (a) follows from the union bound. We then bound the second term as
\begin{align}
  & \E{\frac{1}{M_1M_2M_0}\sum_{w_1',w_1,w_2,w_0}\sum_{y_2} W^{\otimes n}_{Y_2|X_1X_2}(y_2^n| x_1^n(w_1,w_0),x_2^n(w_2,w_0))\indic{\mathcal{E}_{w_1'}}}\\
  &=\frac{1}{M_1M_2M_0}\sum_{w_1,w_2,w_0,w_1'\neq w_1}\sum_{u^n}\sum_{x_1^n}\sum_{x_2^n}\sum_{x_1'^n}\nonumber\\
  &\phantom{=============}\sum_{y_2^n}P_{U}^{\otimes n}(u^n)P_{X_1|U}^{\otimes n}(x_1^n|u^n)P_{X_2|U}^{\otimes n}(x_2^n|u^n)P_{X_1|U}^{\otimes n}(x_1'^n|u^n)\nonumber\\
  &\phantom{=============}\times W_{Y_2|X_1X_2}^{\otimes n}(y_2^n|x_1^n,x_2^n)\indic{\log\frac{W_{Y_2|X_1X_2}^{\otimes n}(y_2^n|x_1'^n,x_2^n)}{P_{Y_2|X_2U}^{\otimes n}(y_2^n|x_2^n,u^n)}\geq \gamma_1}\\
  &= \frac{1}{M_1M_2M_0}\sum_{w_1,w_2,w_0,w_1'\neq w_1}\sum_{u^n}\sum_{x_2^n}\sum_{x_1'^n}\sum_{y_2^n}P_{U}^{\otimes n}(u^n)P_{X_2|U}^{\otimes n}(x_2^n|u^n)P_{X_1|U}^{\otimes n}(x_1'^n|u^n)\nonumber\\
  &\phantom{==}\times  P_{Y_2|X_2U}^{\otimes n}(y_2^n|x_2^n,u^n)\indic{\log\frac{W_{Y_2|X_1X_2}^{\otimes n}(y_2^n|x_1'^n,x_2^n)}{P_{Y_2|X_2U}^{\otimes n}(y_2^n|x_2^n,u^n)}\geq \gamma_1}\\
  &\leq \frac{1}{M_1M_2M_0}\sum_{w_1,w_2,w_0,w_1'\neq w_1}\sum_{y_2^n}P_{U}^{\otimes n}(u^n)P_{X_2|U}^{\otimes n}(x_2^n|u^n)P_{X_1|U}^{\otimes n}(x_1'^n|u^n) W_{Y_2|X_1X_2}^{\otimes n}(y_2^n|x_1'^n,x_2^n)e^{-\gamma_1}\\
  &= (M_1-1)e^{-\gamma_1}\\
  &\leq M_1e^{-\gamma_1}.\label{eq:message_bound}
\end{align}
For the first term in \eqref{eq:Pe_b}, we have
\begin{align}
  &\P[W_{Y_2|X_1X_2}^{\otimes n}P_U^{\otimes n}P_{X_1|U}^{\otimes n}P_{X_2|U}^{\otimes n}]{\log\frac{W_{Y_2|X_1,X_2}^{\otimes n}(Y_2^n|X_1^n,X_2^n)}{P_{Y_2|X_2,U}^{\otimes n}(Y_2^n|X_2^n,U^n)}<\gamma_1}\nonumber\\
  &= \P[W_{Y_2|X_1X_2}^{\otimes n}P_U^{\otimes n}P_{X_1|U}^{\otimes n}P_{X_2|U}^{\otimes n}]{\sum_{i=1}^n\log\frac{W_{Y_2|X_1,X_2}(Y_{2,i}|X_{1,i},X_{2,i})}{P_{Y_2|X_2,U}(Y_{2,i}|X_{2,i},U_i)}<\gamma_1}\\
  &= \P{\sum_{i=1}^nI_{1,i}>n\mu\I{X_1}{Y_2|X_2,U}},
\end{align}
where we have used the definition of $\gamma_1=(1-\mu)n\I{X_1}{Y_2|X_2,U}$ and defined the \ac{iid} random variables $\{I_{1,i}\}_{i=1}^n$  distributed according to the zero-mean random variable $I_1 \eqdef \I{X_1}{Y_2|X_2U}-\log\frac{W_{Y_2|X_1X_2}(Y_2|X_1X_2)}{P_{Y_2|X_2U}(Y_2|X_2U)}$. In order to apply Bernstein's inequality (recalled in Appendix~\ref{sec:bernst-ineq}) we need to evaluate the variance of $I_1$.
\begin{align}
  &\E{\log^2\frac{W_{Y_2|X_1X_2}(Y_2|X_1X_2)}{P_{Y_2|X_2U}(Y_2|X_2U)}}\nonumber\\
  &= \sum_{u,x_1,x_2,y_2} P_U(u)P_{X_1|U}(x_1|u)P_{X_2|U}(x_2|u)P^{(2)}_{x_1x_2}(y_2)\times\log^2\frac{P^{(2)}_{x_1x_2}(y_2)}{\sum_{x_1'}P_{X_1|U}(x_1'|u)P^{(2)}_{x_1'x_2}(y_2)}\\
    &= q_1n^{-\frac{1}{4}}\sum_{x_1,y_2}P_{X_1|U}(x_1|1)P^{(2)}_{x_10}(y_2)\left(\log^2\frac{P^{(2)}_{x_10}(y_2)}{P^{(2)}_{00}(y_2)}\right.\\
    &\phantom{==}-2\log\left(\frac{P^{(2)}_{x_10}}{P^{(2)}_{00}(y_2)}\right)\log\left(1+p_1n^{-\frac{1}{4}}\frac{P^{(2)}_{10}(y_2)-P^{(2)}_{00}(y_2)}{P^{(2)}_{00}(y_2)}\right)\nonumber\\
    &\phantom{==}\left.+\log^2\left(1+p_1n^{-\frac{1}{4}}\frac{P^{(2)}_{10}(y_2)-P^{(2)}_{00}(y_2)}{P^{(2)}_{00}(y_2)}\right)\right)\\
    &= \calO\left(n^{-\frac{1}{2}}\right).
\end{align}
Since $P_{Y_2|X_1X_2}\ll P_{Y_2|X_2U}$, there exists $c>0$ such that $\abs{I_1}\leq c$, and we can apply Bernstein's inequality
\begin{align}
    \P{\sum_{i=1}^nI_{1,i}\geq\mu n\I{X_1}{Y_2|X_2U}} &\leq \exp\left(-\frac{n^2\mu^2\I{X_1}{Y_2|X_2U}^2}{\sum_{i=1}^n\Var{I_{1,i}}+\frac{1}{3}cn\mu \I{X_1}{Y_2|X_1U}}\right)\\
    &= \exp\left(-\frac{\calO\left(n^{1}\right)}{\calO\left(n^{\frac{1}{2}}\right)}\right)\\
    &\leq \exp\left(-\zeta_1n^{\frac{1}{2}}\right),
\end{align}
for an appropriate constant $\zeta_1>0$.

From~\eqref{eq:message_bound}, if 
\begin{align}
    &\log M_1<(1-\mu)n\I{X_1}{Y_2|X_2,U} = q_1p_1n^{\frac{1}{2}}\D{P^{(2)}_{10}}{P^{(2)}_{00}}+\calO\left(n^{\frac{1}{4}}\right),\\
    &\log M_2<(1-\mu)n\I{X_2}{Y_1|X_1,U} = q_2p_2n^{\frac{1}{2}}\D{P^{(1)}_{01}}{P^{(1)}_{00}}+\calO\left(n^{\frac{1}{4}}\right),
\end{align}
then $\E{P_e^{(n)}}\leq\exp\left(-\xi n^{\frac{1}{2}}\right)$ for an appropriate constant $\xi>0$.

\section{Proof of Lemma ~\ref{lm:resolvability2}}\label{appendix:proof_resolvability2}
For $\bftau\eqdef\set{\tau_i}_{i=1}^4$, we first define the set
\begin{equation}
  \calB^{(n)}_\tau = \bigcap_{i=1}^4\calB^{(n)}_{\tau_i},
\end{equation}
where
\begin{align}
  &\calB^{(n)}_{\tau_1} = \left\{(x_1^n,x_2^n,u^n,z^n):\log\frac{W_{Z|X_1,X_2}^{\otimes n}(z^n|z_1^n,x_2^n)}{Q_Z^{\otimes n}(z^n)}\leq\tau_1\right\},\\
  &\calB^{(n)}_{\tau_2} = \left\{(x_1^n,x_2^n,u^n,z^n):\log\frac{P_{Z|X_1,U}^{\otimes n}(z^n|x_1^n,u^n)}{Q_Z^{\otimes n}(z^n)}\leq\tau_2\right\},\\
  &\calB^{(n)}_{\tau_3} = \left\{(x_1^n,x_2^n,u^n,z^n):\log\frac{P_{Z|X_2,U}^{\otimes n}(z^n|x_2^n,u^n)}{Q_Z^{\otimes n}(z^n)}\leq\tau_3\right\},\\
  &\calB^{(n)}_{\tau_4} = \left\{(x_1^n,x_2^n,u^n,z^n):\log\frac{P_{Z|U}^{\otimes n}(z^n|u^n)}{Q_Z^{\otimes n}(z^n)}\leq\tau_4\right\},
\end{align}
where $\tau_1~\eqdef(1+\mu)n\I{X_1,X_2}{Z}$, $\tau_2~\eqdef(1+\mu)n\I{X_1,U}{Z}$, $\tau_3~\eqdef(1+\mu)n\I{X_2,U}{Z}$, $\tau_4~\eqdef(1+\mu)n\I{U}{Z}$.

Taking the expectation of the relative entropy $\D{P_{Z^n,W_1^s,W_2^s}}{Q_{Z}^{\otimes n}P_{W_1^s}P_{W_2^s}}$ with respect to the random codebook, we obtain
\begin{align}
    &\E[C_n]{\D{P_{Z^n,W_1^s,W_2^s}}{Q_{Z}^{\otimes n}P_{W_1^s}P_{W_2^s}}}\nonumber\\
    &\stackrel{(a)}{=} \sum_{z^n}\mathbb{E}_{U^n(1)X_1^n(1,1,1)X_2^n(1,1,1)}\left[{W^{\otimes n}_{Z|X_1,X_2}(z^n|X_1^n(1,1,1)x_2^n(1,1,1))} \right.\nonumber\\
    & \times\left. \E[C_n\setminus\{U^n(1)X_1^n(1,1,1)X_2^n(1,1,1)\}]{\log{\frac{1}{M_1^pM_2^pM_0}\sum_{w_1^{p'},w_2^{p'},w_0'}}\frac{W^{\otimes n}_{Z|X_1,X_2}(z^n|X_1^n(w_1^{p'},1,w_0')X_2^n(w_2^{p'},1,w_0'))}{Q_{Z}^{\otimes n}(z^n)}}\right]\\
      & \stackrel{(b)}{\leq}\sum_{z^n}\sum_{u^n(1)}\sum_{x_1^n(1,1,1)}\sum_{x_2^n(1,1,1)}W^{\otimes n}_{Z|X_1,X_2}(z^n|x_1^n(1,1,1)x_2^n(1,1,1))\nonumber\\
  &\phantom{==========}P_{U}^{\otimes n}(u^n(1))P_{X_1|U}^{\otimes n}(x_1^n(1,1,1)|u^n(1))P_{X_2|U}^{\otimes n}(x_2^n(1,1,1)|u^n(1))\nonumber\\
    & \times \log{\E[C_n\setminus\{u^n(1)x_1^n(1,1,1)x_2^n(1,1,1)\}]{\frac{1}{M_1^pM_2^pM_0}\sum_{w_1^{p'},w_2^{p'},w_0}\frac{W^{\otimes n}_{Z|X_1,X_2}(z^n|X_1^n(w_1^{p'},w_1^s,w_0')X_2^n(w_2^{p'},w_2^s,w_0'))}{Q_{Z}^{\otimes n}(z^n)}}} \label{eq:concavity_bound}
\end{align}
where $(a)$ follows by expanding $P_{Z^n,W_1^s,W_2^s}$ and using the symmetry of the random code construction; $(b)$ follows by explicitly writing $\mathbb{E}_{U^n(1)X_1^n(1,1,1)X_2^n(1,1,1)}$ and from the concavity of $\log\left(\cdot\right)$. The $\log$ term can be expanded and simplified as
\begin{align}
  &  \log{\left[\frac{1}{M_1^pM_2^pM_0}\left(\frac{W^{\otimes n}_{Z|X_1,X_2}(z^n|x_1^n(1,1,1),x_2(1,1,1))}{Q_{Z}^{\otimes n}(z^n)}\right.\right.}\nonumber\\
  &+\sum_{{w_1^{p'} \neq 1}} \E[X_1^n(w_1^{p'},1,1)]{\frac{W^{\otimes n}_{Z|X_1,X_2}(z^n|X_1^n(w_1^{p'},1,1),x_2^n(1,1,1))}{Q_{Z}^{\otimes n}(z^n)}\Bigg|U^n=u^n(1)} \nonumber\\
  &+\sum_{{w_2^{p'} \neq 1}} \E[X_2^n(w_2^{p'},1,1)]{\frac{W^{\otimes n}_{Z|X_1,X_2}(z^n|x_1^n(1,1,1),X_2^n(w_2^{p'},1,1))}{Q_{Z}^{\otimes n}(z^n)}\Bigg|U^n=u^n(1)}\nonumber\\
  &+\sum_{{w_1^{p'} \neq w_1^p,w_2^{p'}\neq w_2^p}}\E[X_1^n(w_1^{p'},1,1),X_2^n(w_2^{p'},1,1)]{\frac{W^{\otimes n}_{Z|X_1,X_2}(z^n|X_1^n(w_1^{p'},1,1),X_2^n(w_2^{p'},1,1))}{Q_{Z}^{\otimes n}(z^n)}\Bigg|U^n=u^n(1)} \nonumber\\
  &\left.\left.+\sum_{{w_1^{p'} \neq w_1^p, w_2^{p'} \neq w_2^p, w_0' \neq w_0}} \E[U^n(w_0'),X_1^n(w_1^{p'},1,w_0'),X_2^n(w_2^{p'},1,w_0')]{\frac{W^{\otimes n}_{Z|X_1,X_2}(z^n|X_1^n(w_1^{p'},1,w_0'),X_2^n(w_2^{p'},1,w_0'))}{Q_{Z}^{\otimes n}(z^n)}}\right)\right]\\
  &= \log\left[\frac{1}{M_1^pM_2^pM_0}\left(\frac{W^{\otimes n}_{Z|X_1,X_2}(z^n|x_1^n(1,1,1),x_2(1,1,1))}{Q_{Z}^{\otimes n}(z^n)}\right.\right.\nonumber\\
  &\phantom{======}+\sum_{{w_1^{p'} \neq 1}}\frac{P_{Z|X_2U}^{\otimes n}(z^n|x_2^n(1,1,1),u^n(1))}{Q_{Z}^{\otimes n}(z^n)}+\sum_{{w_2^{p'} \neq 1}}\frac{P_{Z|X_1U}^{\otimes n}(z^n|x_1^n(1,1,1),u^n(1))}{Q_{Z}^{\otimes n}(z^n)}\nonumber\\
  &\phantom{==============}\left.\left.+\sum_{{w_1^{p'} \neq w_1^p, w_2^{p'} \neq w_2^p}}\frac{P_{Z|U}^{\otimes n}(z^n|u^n(1))}{Q_{Z}^{\otimes n}(z^n)}+\sum_{{w_1^{p'} \neq w_1^p, w_2^{p'} \neq w_2^p, w_0' \neq w_0}}1\right)\right].
\end{align}
Hence, recognizing that $x_1^n(1,1,1)$, $x_2^n(1,1,1)$ and $u^n(1)$ are merely indices, the inequality~(\ref{eq:concavity_bound}) takes the simplified form
\begin{align}
  &\E[C_n]{\D{P_{Z^n,W_1^s,W_2^s}}{Q_{Z}^{\otimes n}P_{W_1^s}P_{W_2^s}}}\nonumber\\
  &\leq \sum_{z^n}\sum_{u^n}\sum_{x_1^n}\sum_{x_2^n}W^{\otimes n}_{Z|X_1,X_2}(z^n|x_1^nx_2^n)P_{U}^{\otimes n}(u^n)P_{X_1|U}^{\otimes n}(x_1^n|u^n)P_{X_2|U}^{\otimes n}(x_2^n|u^n)\nonumber\\
  &\phantom{==}\times\log\left[\frac{1}{M_1^pM_2^pM_0}\frac{W^{\otimes n}_{Z|X_1,X_2}(z^n|x_1^n,x_2)}{Q_{Z}^{\otimes n}(z^n)}+\frac{1}{M_2^pM_0}\frac{P_{Z|X_2U}^{\otimes n}(z^n|x_2^n,u^n)}{Q_{Z}^{\otimes n}(z^n)}\right.\nonumber\\
    &\phantom{===============}\left.+\frac{1}{M_1^pM_0}\frac{P_{Z|X_1U}^{\otimes n}(z^n|x_1^n,u^n)}{Q_{Z}^{\otimes n}(z^n)}+\frac{1}{M_0}\frac{P_{Z|U}^{\otimes n}(z^n|u^n)}{Q_{Z}^{\otimes n}(z^n)}+1 \right].\label{eq:simplified_divergence_bound}
\end{align}
If $(u^n,x_1^n,x_2^n,z^n)\in\calB_{\mathbf{\tau}}^{(n)}$, the $\log$ term can be upper bounded by
\begin{multline}
  \log\left[\frac{e^{\tau_1}}{M_1^pM_2^pM_0}+\frac{e^{\tau_2}}{M_2^pM_0}+\frac{e^{\tau_3}}{M_1^pM_0}+\frac{e^{\tau_4}}{M_0}+1\right]\\ \leq \frac{e^{\tau_1}}{M_1^pM_2^pM_0}+\frac{e^{\tau_2}}{M_2^pM_0}+\frac{e^{\tau_3}}{M_1^pM_0}+\frac{e^{\tau_4}}{M_0}.\label{eq:bound_log_1}
\end{multline}
If $(u^n,x_1^n,x_2^n,z^n)\notin\calB_{\mathbf{\tau}}^{(n)}$, define $\nu_\textnormal{min}\eqdef\min_{z:Q_{00}(z)>0}Q_{00}(z)$ and upper bound the $\log$ term as
\begin{align}
&\log\frac{1}{Q_{Z}^{\otimes n}(z^n)}+\log\left[\frac{1}{M_1^pM_2^pM_0}W^{\otimes n}_{Z|X_1,X_2}(z^n|x_1^n,x_2^n)+\frac{1}{M_2^pM_0}P_{Z|X_2U}^{\otimes n}(z^n|x_2^n,u^n)\right.\nonumber\\
  &\phantom{==============}\left.\left.+\frac{1}{M_1^pM_0}P_{Z|X_1U}^{\otimes n}(z^n|x_1^n,u^n)+\frac{1}{M_0}P_{Z|U}^{\otimes n}(z^n|u^n)+Q_{Z}^{\otimes n}(z^n)\right)\right]\\
  &\overset{(a)}{\leq} \left(n\log\frac{1}{(1-n^{-1}(q_1p_1+q_2p_2))\nu_\textnormal{min}}+\log5\right)\\
  &\leq n\log\left(\frac{5}{(1-n^{-1}(q_1p_1+q_2p_2))\nu_\textnormal{min}}\right),\label{eq:bound_log_2}
\end{align}
where $(a)$ follows from upper bounding the terms in the second $\log$ by 1 and using the definition of $\nu_\textnormal{min}$. Substituting~(\ref{eq:bound_log_1}) and~(\ref{eq:bound_log_1}) into~(\ref{eq:simplified_divergence_bound}), we obtain
\begin{align}
  &\E[C_n]{\D{P_{Z^n,W_1^s,W_2^s}}{Q_{Z}^{\otimes n}P_{W_1^s}P_{W_2^s}}}\nonumber\\
  &\leq \P{(\calB_{\mathbf{\tau}}^{(n)})^c} n\log\left(\frac{5}{(1-n^{-1}(q_1p_1+q_2p_2))\nu_\textnormal{min}}\right) + \frac{e^{\tau_1}}{M_1^pM_2^pM_0}+\frac{e^{\tau_2}}{M_2^pM_0}+\frac{e^{\tau_3}}{M_1^pM_0}+\frac{e^{\tau_4}}{M_0}.\label{eq:final_divergence_bound}
\end{align}
Based on the definition of $\bftau$, the last four terms of~(\ref{eq:final_divergence_bound}) vanish provided
\begin{align}
  \log M_1^pM_2^pM_0 &>(1+\mu)n\I{X_1,X_2}{Z},\\
  \log M_2^pM_0&>(1+\mu)n\I{X_2,U}{Z},\\
  \log M_1^pM_0 &>(1+\mu)n\I{X_1,U}{Z},\\
  \log M_0 &> (1+\mu)n\I{U}{Z}.
\end{align}
Using Lemma~\ref{lm:D_localTS2}, these conditions become
\begin{align}
    &\log M_1^pM_2^pM_0 > (1+\mu)n^{\frac{1}{2}}\left(q_1p_1\D{Q_{10}}{Q_{00}}+q_2p_2\D{Q_{01}}{Q_{00}}\right),\label{eq:res_r_cond1}\\
    &\log M_2^pM_0 >  (1+\mu)n^{\frac{1}{2}}q_2p_2\D{Q_{01}}{Q_{00}},\\
    &\log M_1^pM_0 >  (1+\mu)n^{\frac{1}{2}}q_1p_1\D{Q_{10}}{Q_{00}},\\
    &\log M_0 > \calO\left(n^{\frac{1}{4}}\right),\label{eq:res_r_cond4}
\end{align}
which are the desired conditions in Lemma~\ref{lm:resolvability2}. The lemma holds if we show that $\P{(\calB_{\mathbf{\tau}}^{(n)})^c}$ decays fast enough with $n$. From the definition of $\calB^{(n)}_\tau$ and the union bound we have
\begin{align}
  \P{(\calB^{(n)}_\tau)^c} &\leq \sum_{i=1}^4\P{(\calB^{(n)}_{\tau_i})^c}.\label{eq:res_bernstein_b}
\end{align} We define the random variables $V_1 = \log\frac{Q_{X_1X_2}(Z)}{Q_Z(Z)}-\I{X_1,X_2}{Z}$, $V_2 = \frac{P_{Z|X_1U}(Z|X_1,U)}{Q_Z(Z)}-\I{X_1,U}{Z}$, $V_3 = \log\frac{P_{Z|X_2U}(Z|X_2,U)}{Q_Z(Z)}-\I{X_2,U}{Z}$ and $V_4 = \frac{P_{Z|U}(Z|U)}{Q_Z(Z)}-\I{U}{Z}$. For the variance of $V_1$,
\begin{align}
    &\E{\log^2\frac{Q_{X_1X_2}(Z)}{Q_Z(Z)}} \\
    &=\sum_{u,x_1,x_2,z}P_U(u)P_{X_1|U}(x_1|u)P_{X_2|U}(x_2|u)Q_{x_1x_2}(z)\log^2\frac{Q_{x_1x_2}(z)}{Q_Z(z)}\\
    &=q_1n^{-\frac{1}{4}}\sum_{x_1,z}P_{X_1|U}(x_1|1)Q_{x_10}(z)\left(\log^2\frac{Q_{x_10}(z)}{Q_{00}(z)}-2\log\frac{Q_{x_10}(z)}{Q_{00}(z)}\log\frac{Q_Z(z)}{Q_{00}(z)}+\log^2\frac{Q_Z(z)}{Q_{00}(z)}\right)\nonumber\\
    &\phantom{==}+q_2n^{-\frac{1}{4}}\sum_{x_2,z}P_{X_2|U}(x_2|2)Q_{0x_2}(z)\left(\log^2\frac{Q_{0x_2}(z)}{Q_{00}(z)}-2\log\frac{Q_{0x_2}(z)}{Q_{00}(z)}\log\frac{Q_Z(z)}{Q_{00}(z)}+\log^2\frac{Q_Z(z)}{Q_{00}(z)}\right)\\
    &=\calO\left(n^{-\frac{1}{2}}\right),\label{eq:var_V1}
\end{align}
since the lowest order terms in two sums are the first terms. For the variance of $V_2$,
\begin{align}
    &\E{\log^2\frac{P_{Z|X_1U}(Z|X_1,U)}{Q_Z(Z)}} =\\ &\sum_{u,x_1,x_2,z}P_U(u)P_{X_1|U}(x_1|u)P_{X_2|U}(x_2|u)Q_{x_1x_2}(z)\log^2\left(\frac{\sum_{x_2'}P_{X_2|U}(x_2'|u)Q_{x_1x_2'}(z)}{Q_Z(z)}\right)\\
    &= q_1n^{-\frac{1}{4}}\sum_{x_1,z}P_{X_1|U}(x_1|1)Q_{x_10}(z)\left(\log^2\frac{Q_{x_10}(z)}{Q_{00}(z)}-2\log\frac{Q_{x_10}(z)}{Q_{00}(z)}\log\frac{Q_Z(z)}{Q_{00}(z)}+\log^2\frac{Q_Z(z)}{Q_{00}(z)}\right)\nonumber\\
    &\phantom{==}+q_2n^{-\frac{1}{4}}\sum_{x_2,z}P_{X_2|U}(x_2|2)Q_{0x_2}\left(\log^2\left(\frac{\sum_{x_2'}P_{X_2|U}(x_2'|2)Q_{0x_2'}(z)}{Q_{00}(z)}\right)\right.\\
    &\phantom{==}\left.-2\log\left(\frac{\sum_{x_2'}P_{X_2|U}(x_2'|2)Q_{0x_2'}(z)}{Q_{00}(z)}\right)\log\frac{Q_Z(z)}{Q_{00}(z)}+\log^2\frac{Q_Z(z)}{Q_{00}(z)}\right)\\
    &=\calO\left(n^{-\frac{1}{2}}\right),\label{eq:var_V2}
\end{align}
since lowest order term is first term in first summation. By symmetry of $X_1$ and $X_2$,
\begin{align}
    \E{\log^2\frac{P_{Z|X_2U}(Z|X_2,U)}{Q_Z(Z)}} = \calO\left(n^{-\frac{1}{2}}\right).\label{eq:var_V3}
\end{align}

For the variance of $V_4$,
\begin{align}
    &\E{\log^2\frac{P_{Z|U}(Z|U)}{Q_Z(Z)}} \\
    &=\sum_{u,x_1,x_2,z}P_U(u)P_{X_1|U}(x_1|u)P_{X_2|U}(x_2|u)Q_{x_1x_2}(z)\log^2\left(\frac{\sum_{x_1',x_2'}P_{X_1|U}(x_1'|u)P_{X_2|U}(x_2'|u)Q_{x_1'x_2'}(z)}{Q_Z(z)}\right)\\
    &= q_1n^{-\frac{1}{4}}\sum_{x_1,z}\left(Q_{00}(z)+p_1n^{-\frac{1}{4}}\left(Q_{10}(z)-Q_{00}(z)\right)\right)\left(\log^2\left(1+p_1n^{-\frac{1}{4}}\frac{Q_{10}(z)-Q_{00}(z)}{Q_{00}(z)}\right)\right.\nonumber\\
    &\phantom{==}\left.-2\log\left(1+p_1n^{-\frac{1}{4}}\frac{Q_{10}(z)-Q_{00}(z)}{Q_{00}(z)}\right)\log\frac{Q_Z(z)}{Q_{00}(z)}+\log^2\frac{Q_Z(z)}{Q_{00}(z)}\right)\nonumber\\
    &\phantom{==}+q_2n^{-\frac{1}{4}}\sum_{x_2,z}\left(Q_{00}(z)+p_2n^{-\frac{1}{4}}\left(Q_{01}(z)-Q_{00}(z)\right)\right)\left(\log^2\left(1+p_2n^{-\frac{1}{4}}\frac{Q_{01}(z)-Q_{00}(z)}{Q_{00}(z)}\right)\right.\nonumber\\
    &\phantom{==}\left.-2\log\left(1+p_2n^{-\frac{1}{4}}\frac{Q_{01}(z)-Q_{00}(z)}{Q_{00}(z)}\right)\log\frac{Q_Z(z)}{Q_{00}(z)}+\log^2\frac{Q_Z(z)}{Q_{00}(z)}\right)\\
    &= \calO\left(n^{-\frac{3}{4}}\right),\label{eq:var_V4}
\end{align}
since $\log^2\left(1+\calO\left(n^{-\frac{1}{4}}\right)\right)=\calO\left(n^{-\frac{1}{2}}\right)$, $\log\left(1+\calO\left(n^{-\frac{1}{4}}\right)\right)=\calO\left(n^{-\frac{1}{4}}\right)$, $\log^2\frac{Q_Z(z)}{Q_{00}(z)}=\calO\left(n^{-\frac{1}{2}}\right)$ and $\log\frac{Q_Z(z)}{Q_{00}(z)}=\calO\left(n^{-\frac{1}{2}}\right)$ by Lemma~\ref{lm:D_localTS2}.

Using \eqref{eq:var_V1}, \eqref{eq:var_V2}, \eqref{eq:var_V3}, \eqref{eq:var_V4}, Lemma~\ref{lm:D_localTS2}, and Bernstein's inequality, we upper bound~\eqref{eq:res_bernstein_b} as
\begin{align} &\exp\left(-\frac{\calO\left(n\right)}{\calO\left(n^{\frac{1}{2}}\right)}\right)+\exp\left(-\frac{\calO\left(n\right)}{\calO\left(n^{\frac{1}{2}}\right)}\right)+\exp\left(-\frac{\calO\left(n\right)}{\calO\left(n^{\frac{1}{2}}\right)}\right)+\exp\left(-\frac{\calO\left(n^{\frac{1}{2}}\right)}{\calO\left(n^{\frac{1}{4}}\right)}\right)\\
    &\leq \sum_{i=1}^3\exp\left(-\zeta_in^{-\frac{1}{2}}\right)+\exp\left(-\zeta_4n^{-\frac{1}{4}}\right)\\
    &\leq \exp\left(-\zeta n^{-\frac{1}{4}}\right),
\end{align}
for an appropriate constant $\zeta>0$. Hence, if \eqref{eq:res_r_cond1}-\eqref{eq:res_r_cond4} hold, we have
\begin{align}
    \E{\E[W_1^sW_2^s]{\D{P_{Z^n|W_1^sW_2^s}}{Q_Z^{\otimes n}}}}\leq \exp\left(-\xi n^{\frac{1}{4}}\right),
\end{align}
for an appropriate constant $\xi>0$ as desired.

\section{Proof of Lemma ~\ref{lm:resolv_concentration2}}\label{appendix:resolv_concent_proof}
Since $\D{\widehat{Q}^{nB}}{Q_{Z}^{\otimes nB}}\leq \exp\left(-\zeta_2 n^\frac{1}{4}\right)$, from Pinsker's inequality we have $\V{\widehat{Q}^{nB}}{Q_{Z}^{\otimes nB}}\leq \frac{1}{2}\exp\left(-\frac{1}{2}\zeta_2 n^\frac{1}{4}\right)$. Then, we write
\begin{align}
  \D{\widehat{Q}^{nB}}{Q_{00}^{\otimes nB}} &= \D{\widehat{Q}^{nB}}{Q_{Z}^{\otimes nB}}+\D{Q_{Z}^{\otimes nB}}{Q_{00}^{\otimes nB}}\nonumber\\
  &+\sum_{z^{nB}}\left(\widehat{Q}^{nB}(z^{nB})-Q_{Z}^{\otimes nB}(z^{nB})\right)\log\frac{Q_{Z}^{\otimes nB}(z^{nB})}{Q_{00}^{\otimes nB}(z^{nB})}.
\end{align}
By rearranging the terms and taking the absolute value we obtain
\begin{align}
  &\abs{\D{\widehat{Q}^{nB}}{Q_{00}^{\otimes nB}}-\D{Q_{Z}^{\otimes nB}}{Q_{00}^{\otimes nB}}}\nonumber\\
  &\leq \D{\widehat{Q}^{nB}}{Q_{Z}^{\otimes nB}}+\abs{\sum_{z^{nB}}\left(\widehat{Q}^{nB}(z^{nB})-Q_{Z}^{\otimes nB}(z^{nB})\right)\frac{Q_{Z}^{\otimes nB}(z^{nB})}{Q_{00}^{\otimes nB}(z^{nB})}}\\
  &\overset{(a)}{\leq} \D{\widehat{Q}^{nB}}{Q_{Z}^{\otimes nB}}+\V{\widehat{Q}^{nB}}{Q_{Z}^{\otimes nB}}\sup_{z^{nB}}\log\frac{Q_{Z}^{\otimes nB}(z^{nB})}{Q_{00}^{\otimes nB}(z^{nB})}\\
  &\leq \exp\left(-\zeta_2n^\frac{1}{4}\right)+\frac{1}{\sqrt{2}}\exp\left(-\frac{1}{2}\zeta_2n^\frac{1}{4}\right)nB\nonumber\\
  &\times\sup_{z}\log\left(1+n^{-\frac{1}{2}}(q_1p_1+q_2p_2)\frac{\lambda Q_{10}(z)+\Bar{\lambda}Q_{01}(z)-Q_{00}(z)}{Q_{00}(z)}\right)\\
  &\leq \exp\left(-\zeta_2n^\frac{1}{4}\right)+\frac{1}{\sqrt{2}}\exp\left(-\frac{1}{2}\zeta_2n^\frac{1}{4}\right)nB\log\left(1+n^{-\frac{1}{2}}\frac{(q_1p_1+q_2p_2)}{\nu_\textnormal{min}}\right)\\
  &\leq \exp\left(-\zeta_2n^\frac{1}{4}\right)+\frac{1}{\sqrt{2}}\exp\left(-\frac{1}{2}\zeta_2n^\frac{1}{4}\right)n^\frac{1}{2}\frac{(q_1p_1+q_2p_2)}{\nu_\textnormal{min}}B\\
  &\leq \exp\left(-\zeta_3n^\frac{1}{4}\right),
\end{align}
where $(a)$ follows from Holder's Inequality and $\zeta_3>0$ is an appropriate constant.

\section{Proof of Lemma ~\ref{lm:D_conv}}\label{appendix:D_conv}
By the definition of relative entropy
\begin{align}
  \D{\sum_{ij}\mu^n_{ij}Q_{ij}}{Q_{00}} &= \sum_z \left(Q_{00}(z)+\sum_{ij}\mu^n_{ij}(Q_{ij}(z)-Q_{00}(z))\right)\nonumber\\
  &\times\log\left(1+\frac{\sum_{ij}\mu^n_{ij}(Q_{ij}(z)-Q_{00}(z))}{Q_{00}(z)}\right).
\end{align}

Since $\forall (i,j)\neq(0,0),\quad \lim_{n\to\infty}\mu_{ij}^n=0$, there exists $n$ large enough such that $\frac{\sum_{ij}\mu_{ij}^n(Q_{ij}(z)-Q_{00}(z))}{Q_{00}(z)}\geq-\frac{1}{2}$ for all $z\in\calZ$. By using the inequality $\log(1+x)>x-\frac{x^2}{2}\quad \forall x>0$ and $\log\left(1+x\right)\geq x-\frac{x^2}{2}+\frac{2x^3}{3}\quad \forall x\in\left[-\frac{1}{2},0\right]$, for $n$ large enough
\begin{align}
  \D{\sum_{ij}\mu^n_{ij}Q_{ij}}{Q_{00}}&\geq \sum_{z}\left(Q_{00}(z)+\sum_{ij}\mu_{ij}^n(Q_{ij}(z)-Q_{00}(z))\right)\nonumber\\
                                       &\phantom{==}\times\left[\frac{\sum_{kl}\mu_{kl}^n\left(Q_{kl}(z)-Q_{00}(z)\right)}{Q_{00}(z)}-\frac{1}{2}\frac{\left(\sum_{kl}\mu_{kl}^n\left(Q_{kl}(z)-Q_{00}(z)\right)\right)^2}{Q_{00}(z)^2}\right]\nonumber\\
                                       &\phantom{==}+\frac{2}{3}\sum_{z:\overline{Q}^n(z)\leq Q_{00}}\left(Q_{00}(z)+\sum_{ij}\mu_{ij}^n(Q_{ij}(z)-Q_{00}(z))\right)\nonumber\\
  &\phantom{==}\times\frac{\left(\sum_{kl}\mu_{kl}^nQ_{kl}(z)-Q_{00}(z)\right)^3}{Q_{00}(z)^3}\\
  &\approx \frac{1}{2}\chi_2\left(\sum_{ij}\mu_{ij}^nQ_{ij}\middle\Vert Q_{00}\right),\label{eq:D_local2}
\end{align}
where the higher order can be eliminated, since $\left(\mu^n_{ij}\right)^3$ terms go to $0$ faster than $\left(\mu^n_{ij}\right)^2$ terms $\forall (i,j)\neq(0,0)$. Thus, the bound on the weights for $n$ large enough becomes
\begin{align*}
  \frac{2\delta}{n} &\geq \chi_2\left(\sum_{ij}\mu^n_{ij}Q_{ij}\Biggl|\Biggl| Q_{00}\right) = \mu_n^2\chi_2\left(\sum_{ij}\rho^n_{ij}Q_{ij}\Biggl|\Biggl| Q_{00}\right),
\end{align*}
where $\rho_{ij}^n\geq0\quad\forall (i,j)\in[0,1]^2,\forall n\in\bbN^*$, $\rho_{00}^n=0,\forall n\in\bbN^*$ and $\sum_{ij}\rho_{ij}^n=1$. The average code distribution is then
\begin{align}
  P_{\Tilde{X}_1\Tilde{X}_2}(k,\ell) &= \left\{\begin{array}{cc}
                                                 1-\mu_n & (k,\ell)=(0,0) \\
                                                 \mu_n\rho_{k\ell}^n & \textnormal{otherwise}
                                               \end{array}\right.,
\end{align}
Then the constraint on the average weight of the code is
\begin{align}
  \mu_n \leq \sqrt{\frac{2\delta}{n\chi_2\left(\sum_{ij}\rho^n_{ij}Q_{ij}\Biggl|\Biggl| Q_{00}\right)}},
\end{align}
for $n$ large enough.

\section{Proof of Lemma ~\ref{lm:conv_r1r2}}\label{appendix:conv_r1r2}
\begin{align}
  \I{X_{1i}}{Y_{2i}|X_{2i}} &= \I{X_{1i},X_{2i}}{Y_{2i}}-\I{X_{2i}}{Y_{2i}}\label{eq:chain_rule}
\end{align}
Looking at the first term,
\begin{align}
    \I{X_{1i},X_{2i}}{Y_{2i}} &= \sum_{x_1,x_2}P_{X_{1i}X_{2i}}(x_1,x_2)\D{P^{(2)}_{x_1x_2}}{P^{(2)}_{\mu^{(n),i}}}\\
                              &= \sum_{x_1,x_2}P_{X_{1i}X_{2i}}(x_1,x_2)\D{P^{(2)}_{x_1x_2}}{P^{(2)}_{00}}-\D{P^{(2)}_{\mu^{(n),i}}}{P^{(2)}_{00}},\label{eq:IX1X2Y2}
\end{align}
where $P^{(2)}_{\mu^{(n),i}}=\sum_{k,\ell}\mu^{(n),i}_{k\ell}P^{(2)}_{k\ell}$. For the second term in \eqref{eq:chain_rule},
\begin{align}
    \I{X_{2i}}{Y_{2i}} &= \sum_{x_2,y_2}\sum_{x_1}P_{X_{1i}X_{2i}}(x_1,x_2)P^{(2)}_{x_1x_2}(y_2)\log\frac{\sum_{x_1'}P_{X_{1i}X_{2i}}(x_1',x_2)P^{(2)}_{x_1'x_2}(y_2)}{\sum_{\Bar{x}_1}P_{X_1iX_{2i}}(\overline{x_1},x_2)P^{(2)}_{\mu^{(n),i}}(y_2)}\\
                       &= \sum_{x_2,y_2}\sum_{x_1}P_{X_{1i}X_{2i}}(x_1,x_2)P^{(2)}_{x_1x_2}(y_2)\log\frac{\sum_{x_1'}P_{X_{1i}X_{2i}}(x_1',x_2)P^{(2)}_{x_1'x_2}(y_2)}{\sum_{\Bar{x}_1}P_{X_1iX_{2i}}(\overline{x_1},x_2)P^{(2)}_{00}(y_2)}\nonumber\\
                       &-\D{P^{(2)}_{\mu^{(n),i}}}{P^{(2)}_{00}}.\label{eq:IX2Y2}
\end{align}
For $x_2=0$,
\begin{align}
    &\sum_{y_2}\left(\left(1-\left(\mu^{(n),i}-\mu^{(n),i}_{10}\right)\right)P^{(2)}_{00}(y_2)+\mu^{(n),i}_{10}(P^{(2)}_{10}(y_2)-P^{(2)}_{00}(y_2)))\right)\nonumber\\
    &\phantom{==}\times\log\left(1+\frac{\mu^{(n),i}_{10}}{\left(1-\left(\mu^{(n),i}-\mu^{(n),i}_{10}\right)\right)}\frac{P^{(2)}_{10}(y_2)-P^{(2)}_{00}(y_2)}{P^{(2)}_{00}(y_2)}\right)\\
    &\overset{(a)}{=} \calO\left(\frac{\left(\mu^{(n),i}_{10}\right)^2}{1-\left(\mu^{(n),i}-\mu^{(n),i}_{10}\right)}\right)\label{eq:order1}\\
    &\overset{(b)}{=} \calO\left(\left(\mu^{(n),i}_{10}\right)^2\right),\label{eq:order2}
\end{align}
where $(a)$ follows from Lemma~\ref{lm:D_localTS1} and $(b)$ follows from Taylor series expansion of the denominator in \eqref{eq:order1}. For $x_2=1$,
\begin{align}
  \sum_{y_2}\left(\sum_{x_1}\mu_{x_11}^{(n),i}P^{(2)}_{x_11}(y_2)\right)\log\frac{\sum_{x_1'}\mu_{x_1'1}^{(n),i}P^{(2)}_{x_1'1}(y_2)}{\left(\sum_{\Bar{x}_1}\mu_{\Bar{x}_11}^{(n),i}\right)P^{(2)}_{00}(y_2)},\label{eq:IX2Y2_1}
\end{align}
when we push summation over $i$ in \eqref{eq:IX2Y2_1},
\begin{align}
&\sum_{i=1}^n\sum_{y_2}\left(\sum_{x_1}\mu_{x_11}^{(n),i}P^{(2)}_{x_11}(y_2)\right)\log\frac{\sum_{x_1'}\mu_{x_1'1}^{(n),i}P^{(2)}_{x_1'1}(y_2)}{\left(\sum_{\Bar{x}_1}\mu_{\Bar{x}_11}^{(n),i}\right)P^{(2)}_{00}(y_2)}\\
  &= \sum_{y_2}\sum_{i=1}^n\left(\sum_{x_1}\mu_{x_11}^{(n),i}P^{(2)}_{x_11}(y_2)\right)\log\frac{\sum_{x_1'}\mu_{x_1'1}^{(n),i}P^{(2)}_{x_1'1}(y_2)}{\left(\sum_{\Bar{x}_1}\mu_{\Bar{x}_11}^{(n),i}\right)P^{(2)}_{00}(y_2)}\\
  &\geq n\sum_{y_2}\left(\sum_{x_1}\mu_{x_11}^{n}P^{(2)}_{x_11}(y_2)\right)\log\frac{\sum_{x_1'}\mu_{x_1'1}^{n}P^{(2)}_{x_1'1}(y_2)}{\left(\sum_{\Bar{x}_1}\mu_{\Bar{x}_11}^{n}\right)P^{(2)}_{00}(y_2)}\\
  &= n\mu_n\left(\sum_i\rho^n_{i1}\right)\D{\frac{\sum_j\rho^n_{j1}P^{(2)}_{j1}}{\sum_k\rho^n_{k1}}}{P^{(2)}_{00}},\label{eq:log-sum}
\end{align}
by log-sum inequality. Hence, by combining \eqref{eq:IX1X2Y2}, \eqref{eq:IX2Y2}, \eqref{eq:order2} and \eqref{eq:log-sum},
\begin{align}
  \sum_{i=1}^n\I{X_{1i}}{Y_{2i}|X_{2i}}\leq& n\mu_n\left(\sum_{ij}\rho_{ij}^n\D{P^{(2)}_{ij}}{P^{(2)}_{00}}-\left(\sum_i\rho_{i1}^n\right)\D{\frac{\sum_j\rho^n_{j1}P^{(2)}_{j1}}{\sum_k\rho^n_{k1}}}{P^{(2)}_{00}}\right)\nonumber\\
  &+\calO\left(n\left(\mu_{10}^n\right)^2\right).\label{eq:I_bound}
\end{align}

\section{Analysis of Time-Sharing Scheme}\label{appendix:TS1}
Consider the random coding argument in Appendix~\ref{appendix:proof_resolvability2}, considering the same $V_4$ but now under Time-Sharing scheme with distrbution defined in Section~\ref{lm:D_localTS1}. By Lemma~\ref{lm:D_localTS1}, $\I{U}{Z}$ is $\calO\left(n^{-1}\right)$. For the variance of $V_4$,
\begin{align}
    \E{\log^2\frac{P_{Z|U}(Z|U)}{Q_Z(Z)}} &= \sum_{u,x_1,x_2,z}P_U(u)P_{X_1|U}(x_1|u)P_{X_2|U}(x_2|u)Q_{x_1x_2}(z)\nonumber\\
    &\times\left(\log^2\frac{\sum_{x_1',x_2'}P_{X_1|U}(x_1'|u)P_{X_2|U}(x_2'|u)Q_{x_1'x_2'}(z)}{Q_{00}(z)}\right.\nonumber\\
    &\left.-2\log\frac{\sum_{x_1',x_2'}P_{X_1|U}(x_1'|u)P_{X_2|U}(x_2'|u)Q_{x_1'x_2'}(z)}{Q_{00}(z)}\log\frac{Q_Z(z)}{Q_{00}(z)}+\log^2\frac{Q_Z(z)}{Q_{00}(z)}\right)\\
    &= p\sum_z\left(Q_{00}(z)+\calO\left(n^{-1}\right)\right)\left(\log^2\left(1+\calO\left(n^{-\frac{1}{2}}\right)\right)\right.\nonumber\\
    &\left.-2\log\left(1+\calO\left(n^{-\frac{1}{2}}\right)\right)\log\left(1+\calO\left(n^{-\frac{1}{2}}\right)\right)+\log^2\left(1+\calO\left(n^{-\frac{1}{2}}\right)\right) \right)\nonumber\\
    &+(1-p)\sum_z\left(Q_{00}(z)+\calO\left(n^{-1}\right)\right)\left(\log^2\left(1+\calO\left(n^{-\frac{1}{2}}\right)\right)\right.\\
    &\left.-2\log\left(1+\calO\left(n^{-\frac{1}{2}}\right)\right)\log\left(1+\calO\left(n^{-\frac{1}{2}}\right)\right)+\log^2\left(1+\calO\left(n^{-\frac{1}{2}}\right)\right) \right)\\
    &= \calO\left(n^{-1}\right).
\end{align}
Hence Bernstein's Inequality becomes,
\begin{align}
    \P{\sum_{i=1}^nV_{4,i}>n\mu\I{U}{Z}}&\leq \exp\left(-\frac{1}{2}\frac{n^2\mu^2\I{U}{Z}^2}{n\Var{V_4}+\frac{1}{3}cn\mu\I{U}{Z}}\right)\\
    &\leq\exp\left(-\calO(1)\right),
\end{align}
where $c$ is such that $\abs{V_{4}}\leq c<\infty$. We are not aware of a result that would allow the concentration of measure required here.

\end{document}

%% file: Acronyms.tex
\acrodef{ACDIS}[ACDIS]{Adaptive Communication Decision and Information Systems}
\acrodef{AEP}{Asymptotic Equipartition Property}
\acrodef{AoA}{Angle of Arrival}
\acrodef{AWGN}{Additive White Gaussian Noise}
\acrodef{AVC}[AVC]{Arbitrarily Varying Channel}
\acrodef{BER}{Bit-Error-Rate}
\acrodef{BEC}{Binary Erasure Channel}
\acrodef{BPSK}{Binary Phase-Shift Keying}
\acrodef{BSC}{Binary Symmetric Channel}
\acrodef{BICM}[BICM]{Bit-Interleaved Coded-Modulation}
\acrodef{CDF}[CDF]{Cumulative Distribution Function}
\acrodef{CGF}[CGF]{Cumulant Generating Function}
\acrodef{CLT}[CLT]{Central Limit Theorem}
\acrodef{cq}[c-q]{Classical-Quantum}
\acrodef{CSI}[CSI]{Channel State Information}
\acrodef{DMC}[DMC]{Discrete Memoryless Channel}
\acrodef{DMS}[DMS]{Discrete Memoryless Source}
\acrodef{ERM}[ERM]{Empirical Risk Minimization}
\acrodef{FER}[FER]{Frame Error Rate}
\acrodef{ICA}[ICA]{Independent Component Analysis}
\acrodef{iid}[i.i.d.]{independent and identically distributed}
\acrodef{IoT}[IoT]{Internet of Things}
\acrodef{KKT}[KKT]{Karush-Kuhn Tucker}
\acrodef{LASSO}[LASSO]{Least Absolute Shrinkage and Selection Operator}
\acrodef{LPD}[LPD]{Low Probability of Detection}
\acrodef{LDPC}[LDPC]{Low-Density Parity-Check}
\acrodef{LLMS}[LLMS]{Linear Least Mean Square}
\acrodef{LMS}[LMS]{Least Mean Square}
\acrodef{MAC}[MAC]{multiple-access channel}
\acrodef{MGF}[MGF]{Moment Generating Function}
\acrodef{MLC}[MLC]{Multi-Level Coding}
\acrodef{MLE}[MLE]{Maximum Likelihood Estimate}
\acrodef{MIMO}[MIMO]{Multiple-Input Multiple-Output}
\acrodef{MISO}{Multiple-Input Single-Output}
\acrodef{MSD}[MSD]{Multi-Stage Decoding}
\acrodef{MMSE}[MMSE]{Minimum Mean-Square Error}
\acrodef{PAC}[PAC]{Probably Approximately Correct}
\acrodef{PCA}[PCA]{Principal Component Analysis}
\acrodef{PDF}[PDF]{Probability Density Function}
\acrodef{PMF}[PMF]{Probability Mass Function}
\acrodef{POVM}[POVM]{Positive Operator-Valued Measure}
\acrodef{PVM}[PVM]{Projection-Valued Measure}
\acrodef{PPM}[PPM]{Pulse Position Modulation}
\acrodef{PSD}{Power Spectral Density}
\acrodef{PSK}{Phase Shift Keying}
\acrodef{QKD}{Quantum Key Distribution}
\acrodef{ROC}{Receiver Operating Characteristic}
\acrodef{CVQKD}{Continuous-Variable \ac{QKD}}
\acrodef{QPSK}{Quadrature Phase-Shift Keying}
\acrodef{RHS}{right-hand side}
\acrodef{RV}{random variable}
\acrodef{SIMO}{Single-Input Multiple-Output}
\acrodef{SNR}{Signal-to-Noise Ratio}
\acrodef{SVM}[SVM]{Support Vector Machine}
\acrodef{TPCP}{Trace-Preserving Completely-Positive}
\acrodef{wrt}[w.r.t.]{with respect to}
\acrodef{WSS}{Wide Sense Stationary}

%% file: CommandsAndMacros.tex
\newcommand{\matA}{\mathbf{A}}
\newcommand{\bfA}{\mathbf{A}}
\newcommand{\calA}{\mathcal{A}}
\newcommand{\bbA}{\mathbb{A}}
\newcommand{\bfa}{\mathbf{a}}
\newcommand{\veca}{\mathbf{a}}
\newcommand{\matB}{\mathbf{B}}
\newcommand{\bfB}{\mathbf{B}}
\newcommand{\calB}{\mathcal{B}}
\newcommand{\bbB}{\mathbb{B}}
\newcommand{\bfb}{\mathbf{b}}
\newcommand{\vecb}{\mathbf{b}}
\newcommand{\matC}{\mathbf{C}}
\newcommand{\bfC}{\mathbf{C}}
\newcommand{\calC}{\mathcal{C}}
\newcommand{\bbC}{\mathbb{C}}
\newcommand{\bfc}{\mathbf{c}}
\newcommand{\vecc}{\mathbf{c}}
\newcommand{\matD}{\mathbf{D}}
\newcommand{\bfD}{\mathbf{D}}
\newcommand{\calD}{\mathcal{D}}
\newcommand{\bbD}{\mathbb{D}}
\newcommand{\bfd}{\mathbf{d}}
\newcommand{\vecd}{\mathbf{d}}
\newcommand{\matE}{\mathbf{E}}
\newcommand{\bfE}{\mathbf{E}}
\newcommand{\calE}{\mathcal{E}}
\newcommand{\bbE}{\mathbb{E}}
\newcommand{\bfe}{\mathbf{e}}
\newcommand{\vece}{\mathbf{e}}
\newcommand{\matF}{\mathbf{F}}
\newcommand{\bfF}{\mathbf{F}}
\newcommand{\calF}{\mathcal{F}}
\newcommand{\bbF}{\mathbb{F}}
\newcommand{\bff}{\mathbf{f}}
\newcommand{\vecf}{\mathbf{f}}
\newcommand{\matG}{\mathbf{G}}
\newcommand{\bfG}{\mathbf{G}}
\newcommand{\calG}{\mathcal{G}}
\newcommand{\bbG}{\mathbb{G}}
\newcommand{\bfg}{\mathbf{g}}
\newcommand{\vecg}{\mathbf{g}}
\newcommand{\matH}{\mathbf{H}}
\newcommand{\bfH}{\mathbf{H}}
\newcommand{\calH}{\mathcal{H}}
\newcommand{\bbH}{\mathbb{H}}
\newcommand{\bfh}{\mathbf{h}}
\newcommand{\vech}{\mathbf{h}}
\newcommand{\matI}{\mathbf{I}}
\newcommand{\bfI}{\mathbf{I}}
\newcommand{\calI}{\mathcal{I}}
\newcommand{\bbI}{\mathbb{I}}
\newcommand{\bfi}{\mathbf{i}}
\newcommand{\veci}{\mathbf{i}}
\newcommand{\matJ}{\mathbf{J}}
\newcommand{\bfJ}{\mathbf{J}}
\newcommand{\calJ}{\mathcal{J}}
\newcommand{\bbJ}{\mathbb{J}}
\newcommand{\bfj}{\mathbf{j}}
\newcommand{\vecj}{\mathbf{j}}
\newcommand{\matK}{\mathbf{K}}
\newcommand{\bfK}{\mathbf{K}}
\newcommand{\calK}{\mathcal{K}}
\newcommand{\bbK}{\mathbb{K}}
\newcommand{\bfk}{\mathbf{k}}
\newcommand{\veck}{\mathbf{k}}
\newcommand{\matL}{\mathbf{L}}
\newcommand{\bfL}{\mathbf{L}}
\newcommand{\calL}{\mathcal{L}}
\newcommand{\bbL}{\mathbb{L}}
\newcommand{\bfl}{\mathbf{l}}
\newcommand{\vecl}{\mathbf{l}}
\newcommand{\matM}{\mathbf{M}}
\newcommand{\bfM}{\mathbf{M}}
\newcommand{\calM}{\mathcal{M}}
\newcommand{\bbM}{\mathbb{M}}
\newcommand{\bfm}{\mathbf{m}}
\newcommand{\vecm}{\mathbf{m}}
\newcommand{\matN}{\mathbf{N}}
\newcommand{\bfN}{\mathbf{N}}
\newcommand{\calN}{\mathcal{N}}
\newcommand{\bbN}{\mathbb{N}}
\newcommand{\bfn}{\mathbf{n}}
\newcommand{\vecn}{\mathbf{n}}
\newcommand{\matO}{\mathbf{O}}
\newcommand{\bfO}{\mathbf{O}}
\newcommand{\calO}{\mathcal{O}}
\newcommand{\bbO}{\mathbb{O}}
\newcommand{\bfo}{\mathbf{o}}
\newcommand{\veco}{\mathbf{o}}
\newcommand{\matP}{\mathbf{P}}
\newcommand{\bfP}{\mathbf{P}}
\newcommand{\calP}{\mathcal{P}}
\newcommand{\bbP}{\mathbb{P}}
\newcommand{\bfp}{\mathbf{p}}
\newcommand{\vecp}{\mathbf{p}}
\newcommand{\matQ}{\mathbf{Q}}
\newcommand{\bfQ}{\mathbf{Q}}
\newcommand{\calQ}{\mathcal{Q}}
\newcommand{\bbQ}{\mathbb{Q}}
\newcommand{\bfq}{\mathbf{q}}
\newcommand{\vecq}{\mathbf{q}}
\newcommand{\matR}{\mathbf{R}}
\newcommand{\bfR}{\mathbf{R}}
\newcommand{\calR}{\mathcal{R}}
\newcommand{\bbR}{\mathbb{R}}
\newcommand{\bfr}{\mathbf{r}}
\newcommand{\vecr}{\mathbf{r}}
\newcommand{\matS}{\mathbf{S}}
\newcommand{\bfS}{\mathbf{S}}
\newcommand{\calS}{\mathcal{S}}
\newcommand{\bbS}{\mathbb{S}}
\newcommand{\bfs}{\mathbf{s}}
\newcommand{\vecs}{\mathbf{s}}
\newcommand{\matT}{\mathbf{T}}
\newcommand{\bfT}{\mathbf{T}}
\newcommand{\calT}{\mathcal{T}}
\newcommand{\bbT}{\mathbb{T}}
\newcommand{\bft}{\mathbf{t}}
\newcommand{\vect}{\mathbf{t}}
\newcommand{\matU}{\mathbf{U}}
\newcommand{\bfU}{\mathbf{U}}
\newcommand{\calU}{\mathcal{U}}
\newcommand{\bbU}{\mathbb{U}}
\newcommand{\bfu}{\mathbf{u}}
\newcommand{\vecu}{\mathbf{u}}
\newcommand{\matV}{\mathbf{V}}
\newcommand{\bfV}{\mathbf{V}}
\newcommand{\calV}{\mathcal{V}}
\newcommand{\bbV}{\mathbb{V}}
\newcommand{\bfv}{\mathbf{v}}
\newcommand{\vecv}{\mathbf{v}}
\newcommand{\matW}{\mathbf{W}}
\newcommand{\bfW}{\mathbf{W}}
\newcommand{\calW}{\mathcal{W}}
\newcommand{\bbW}{\mathbb{W}}
\newcommand{\bfw}{\mathbf{w}}
\newcommand{\vecw}{\mathbf{w}}
\newcommand{\matX}{\mathbf{X}}
\newcommand{\bfX}{\mathbf{X}}
\newcommand{\calX}{\mathcal{X}}
\newcommand{\bbX}{\mathbb{X}}
\newcommand{\bfx}{\mathbf{x}}
\newcommand{\vecx}{\mathbf{x}}
\newcommand{\matY}{\mathbf{Y}}
\newcommand{\bfY}{\mathbf{Y}}
\newcommand{\calY}{\mathcal{Y}}
\newcommand{\bbY}{\mathbb{Y}}
\newcommand{\bfy}{\mathbf{y}}
\newcommand{\vecy}{\mathbf{y}}
\newcommand{\matZ}{\mathbf{Z}}
\newcommand{\bfZ}{\mathbf{Z}}
\newcommand{\calZ}{\mathcal{Z}}
\newcommand{\bbZ}{\mathbb{Z}}
\newcommand{\bfz}{\mathbf{z}}
\newcommand{\vecz}{\mathbf{z}}
\newcommand{\bfalpha}{\boldsymbol{\alpha}}
\newcommand{\vecalpha}{\boldsymbol{\alpha}}
\newcommand{\bfbeta}{\boldsymbol{\beta}}
\newcommand{\vecbeta}{\boldsymbol{\beta}}
\newcommand{\bfgamma}{\boldsymbol{\gamma}}
\newcommand{\vecgamma}{\boldsymbol{\gamma}}
\newcommand{\bfdelta}{\boldsymbol{\delta}}
\newcommand{\vecdelta}{\boldsymbol{\delta}}
\newcommand{\bfepsilon}{\boldsymbol{\epsilon}}
\newcommand{\vecepsilon}{\boldsymbol{\epsilon}}
\newcommand{\bfzeta}{\boldsymbol{\zeta}}
\newcommand{\veczeta}{\boldsymbol{\zeta}}
\newcommand{\bfeta}{\boldsymbol{\eta}}
\newcommand{\veceta}{\boldsymbol{\eta}}
\newcommand{\bftheta}{\boldsymbol{\theta}}
\newcommand{\vectheta}{\boldsymbol{\theta}}
\newcommand{\bfiota}{\boldsymbol{\iota}}
\newcommand{\veciota}{\boldsymbol{\iota}}
\newcommand{\bfkappa}{\boldsymbol{\kappa}}
\newcommand{\veckappa}{\boldsymbol{\kappa}}
\newcommand{\bflamda}{\boldsymbol{\lamda}}
\newcommand{\veclamda}{\boldsymbol{\lamda}}
\newcommand{\bfmu}{\boldsymbol{\mu}}
\newcommand{\vecmu}{\boldsymbol{\mu}}
\newcommand{\bfnu}{\boldsymbol{\nu}}
\newcommand{\vecnu}{\boldsymbol{\nu}}
\newcommand{\bfxi}{\boldsymbol{\xi}}
\newcommand{\vecxi}{\boldsymbol{\xi}}
\newcommand{\bfomicron}{\boldsymbol{\omicron}}
\newcommand{\vecomicron}{\boldsymbol{\omicron}}
\newcommand{\bfpi}{\boldsymbol{\pi}}
\newcommand{\vecpi}{\boldsymbol{\pi}}
\newcommand{\bfrho}{\boldsymbol{\rho}}
\newcommand{\vecrho}{\boldsymbol{\rho}}
\newcommand{\bfsigma}{\boldsymbol{\sigma}}
\newcommand{\vecsigma}{\boldsymbol{\sigma}}
\newcommand{\bftau}{\boldsymbol{\tau}}
\newcommand{\vectau}{\boldsymbol{\tau}}
\newcommand{\bfupsilon}{\boldsymbol{\upsilon}}
\newcommand{\vecupsilon}{\boldsymbol{\upsilon}}
\newcommand{\bfphi}{\boldsymbol{\phi}}
\newcommand{\vecphi}{\boldsymbol{\phi}}
\newcommand{\bfchi}{\boldsymbol{\chi}}
\newcommand{\vecchi}{\boldsymbol{\chi}}
\newcommand{\bfpsi}{\boldsymbol{\psi}}
\newcommand{\vecpsi}{\boldsymbol{\psi}}
\newcommand{\bfomega}{\boldsymbol{\omega}}
\newcommand{\vecomega}{\boldsymbol{\omega}}
\newcommand{\bfAlpha}{\boldsymbol{\Alpha}}
\newcommand{\matAlpha}{\boldsymbol{\Alpha}}
\newcommand{\bfBeta}{\boldsymbol{\Beta}}
\newcommand{\matBeta}{\boldsymbol{\Beta}}
\newcommand{\bfGamma}{\boldsymbol{\Gamma}}
\newcommand{\matGamma}{\boldsymbol{\Gamma}}
\newcommand{\bfDelta}{\boldsymbol{\Delta}}
\newcommand{\matDelta}{\boldsymbol{\Delta}}
\newcommand{\bfEpsilon}{\boldsymbol{\Epsilon}}
\newcommand{\matEpsilon}{\boldsymbol{\Epsilon}}
\newcommand{\bfZeta}{\boldsymbol{\Zeta}}
\newcommand{\matZeta}{\boldsymbol{\Zeta}}
\newcommand{\bfEta}{\boldsymbol{\Eta}}
\newcommand{\matEta}{\boldsymbol{\Eta}}
\newcommand{\bfTheta}{\boldsymbol{\Theta}}
\newcommand{\matTheta}{\boldsymbol{\Theta}}
\newcommand{\bfIota}{\boldsymbol{\Iota}}
\newcommand{\matIota}{\boldsymbol{\Iota}}
\newcommand{\bfKappa}{\boldsymbol{\Kappa}}
\newcommand{\matKappa}{\boldsymbol{\Kappa}}
\newcommand{\bfLamda}{\boldsymbol{\Lamda}}
\newcommand{\matLamda}{\boldsymbol{\Lamda}}
\newcommand{\bfMu}{\boldsymbol{\Mu}}
\newcommand{\matMu}{\boldsymbol{\Mu}}
\newcommand{\bfNu}{\boldsymbol{\Nu}}
\newcommand{\matNu}{\boldsymbol{\Nu}}
\newcommand{\bfXi}{\boldsymbol{\Xi}}
\newcommand{\matXi}{\boldsymbol{\Xi}}
\newcommand{\bfOmicron}{\boldsymbol{\Omicron}}
\newcommand{\matOmicron}{\boldsymbol{\Omicron}}
\newcommand{\bfPi}{\boldsymbol{\Pi}}
\newcommand{\matPi}{\boldsymbol{\Pi}}
\newcommand{\bfRho}{\boldsymbol{\Rho}}
\newcommand{\matRho}{\boldsymbol{\Rho}}
\newcommand{\bfSigma}{\boldsymbol{\Sigma}}
\newcommand{\matSigma}{\boldsymbol{\Sigma}}
\newcommand{\bfTau}{\boldsymbol{\Tau}}
\newcommand{\matTau}{\boldsymbol{\Tau}}
\newcommand{\bfUpsilon}{\boldsymbol{\Upsilon}}
\newcommand{\matUpsilon}{\boldsymbol{\Upsilon}}
\newcommand{\bfPhi}{\boldsymbol{\Phi}}
\newcommand{\matPhi}{\boldsymbol{\Phi}}
\newcommand{\bfChi}{\boldsymbol{\Chi}}
\newcommand{\matChi}{\boldsymbol{\Chi}}
\newcommand{\bfPsi}{\boldsymbol{\Psi}}
\newcommand{\matPsi}{\boldsymbol{\Psi}}
\newcommand{\bfOmega}{\boldsymbol{\Omega}}
\newcommand{\matOmega}{\boldsymbol{\Omega}}
\newcommand{\bfzero}{\boldsymbol{0}}

\newcommand{\card}[1]{{|{#1}|}}
\newcommand{\abs}[1]{{\left|{#1}\right|}}
\newcommand{\sgn}[1]{{\text{sgn}\left({#1}\right)}}
\newcommand{\norm}[2][]{{\left\Vert{#2}\right\Vert}_{#1}}
\newcommand{\inprod}[3][]{{\left\langle{#2},{#3}\right\rangle}_{#1}}
\newcommand{\dotp}[3][]{{\left\langle{#2},{#3}\right\rangle}_{#1}}
\newcommand{\set}[1]{{\{#1\}}}
\newcommand{\trace}[1]{\text{trace}\left(#1\right)}
\newcommand{\argmin}{\mathop{\text{argmin}}}
\newcommand{\argmax}{\mathop{\text{argmax}}}
\newcommand{\intseq}[2]{[{#1};{#2}]}
\newcommand{\eqdef}{\triangleq}
\newcommand{\E}[2][]{{\mathbb{E}_{#1}\left[#2\right]}}
\newcommand{\Var}[1]{{\text{Var}\left(#1\right)}}
\newcommand{\Cov}[1]{{\text{Cov}\left(#1\right)}}
\renewcommand{\P}[2][]{{\mathbb{P}_{#1}\left(#2\right)}}
\newcommand{\indic}[1]{{\mathbf{1}\left\{#1\right\}}}
\newcommand{\mat}[2]{\left[\begin{array}{#1}#2\end{array}\right]}
\newcommand{\bra}[1]{{\left\langle{#1}\right\vert}}
\newcommand{\ket}[1]{{\left\vert{#1}\right\rangle}}

\newcommand{\D}[3][]{{\mathbb{D}_{#1}}\left(#2\,\middle\Vert\,#3\right)} 
\newcommand{\V}[3][]{{\mathbb{V}_{#1}}\!\left(#2,#3\right)} 
\newcommand{\I}[2]{{{\mathbb{I}}\left(#1;#2\right)}} 
\newcommand{\avgI}[1]{{{\mathbb{I}}\!\left(#1\right)}} 
\newcommand{\avgH}[1]{{\mathbb{H}}\!\left(#1\right)}
\newcommand{\h}[1]{{{h}}\left(#1\right)} 
\newcommand{\Hb}[1]{{h_b}\left(#1\right)} 
\newcommand{\He}[1]{{{\mathbb{H}}\!\left(#1\right)}} 

\newcommand{\GF}[1]{\ensuremath{\mbox{GF}(#1)}} 
\newcommand{\F}[1]{\ensuremath{\mathbb{F}_{#1}}} 
\newcommand{\dH}[1]{\ensuremath{\mbox{d}_{\mbox{\tiny H}}(#1)}} 
\newcommand{\wt}[1]{\ensuremath{\mbox{wt}(#1)}} 

\newcommand{\dd}{\text{\textnormal{d}}} 
\renewcommand{\leq}{\leqslant} 
\renewcommand{\geq}{\geqslant} 

\newcommand{\tr}[2][]{\ensuremath{\text{\textnormal{tr}}_{#1}\left(#2\right)}}  
\newcommand{\rk}[1]{\ensuremath{\text{\textnormal{rank}}\left(#1\right)}}  
\renewcommand{\det}[1]{{\left|{#1}\right|}}                              
\renewcommand{\ker}[1]{\ensuremath{\text{\textnormal{Ker}}\left(#1\right)}}  

%% file: Two_Way_Covert_Single.bbl
\begin{thebibliography}{10}
\providecommand{\url}[1]{#1}
\csname url@samestyle\endcsname
\providecommand{\newblock}{\relax}
\providecommand{\bibinfo}[2]{#2}
\providecommand{\BIBentrySTDinterwordspacing}{\spaceskip=0pt\relax}
\providecommand{\BIBentryALTinterwordstretchfactor}{4}
\providecommand{\BIBentryALTinterwordspacing}{\spaceskip=\fontdimen2\font plus
\BIBentryALTinterwordstretchfactor\fontdimen3\font minus
  \fontdimen4\font\relax}
\providecommand{\BIBforeignlanguage}[2]{{%
\expandafter\ifx\csname l@#1\endcsname\relax
\typeout{** WARNING: IEEEtran.bst: No hyphenation pattern has been}%
\typeout{** loaded for the language `#1'. Using the pattern for}%
\typeout{** the default language instead.}%
\else
\language=\csname l@#1\endcsname
\fi
#2}}
\providecommand{\BIBdecl}{\relax}
\BIBdecl

\bibitem{Fridrich2009}
J.~Fridrich, \emph{Steganography in Digital Media: Principles, Algorithms, and
  Applications}.\hskip 1em plus 0.5em minus 0.4em\relax Cambridge University
  Press, Nov. 2009.

\bibitem{Bash2013}
B.~Bash, D.~Goeckel, and D.~Towsley, ``Limits of reliable communication with
  low probability of detection on {AWGN} channels,'' \emph{{IEEE} {J}ournal on
  {S}elected {A}reas in {C}ommunications}, vol.~31, no.~9, pp. 1921--1930, Sep.
  2013.

\bibitem{Wang2016b}
L.~Wang, G.~W. Wornell, and L.~Zheng, ``Fundamental limits of communication
  with low probability of detection,'' \emph{IEEE Transactions on Information
  Theory}, vol.~62, no.~6, pp. 3493--3503, Jun. 2016.

\bibitem{Bloch2015b}
M.~R. Bloch, ``Covert communication over noisy channels: A resolvability
  perspective,'' \emph{IEEE Transactions on Information Theory}, vol.~62,
  no.~5, pp. 2334--2354, May 2016.

\bibitem{Tahmasbi2017}
M.~Tahmasbi and M.~R. Bloch, ``First and second order asymptotics in covert
  communication,'' \emph{IEEE Transactions on Information Theory}, vol.~65,
  no.~4, pp. 2190 --2212, Apr. 2019.

\bibitem{Wang2016c}
L.~Wang, ``Optimal throughput for covert communication over a classical-quantum
  channel,'' in \emph{Proc. of IEEE Information Theory Workshop}, Cambridge,
  UK, Sep. 2016, pp. 364--368.

\bibitem{Sheikholeslami2016}
A.~Sheikholeslami, B.~A. Bash, D.~Towsley, D.~Goeckel, and S.~Guha, ``Covert
  communication over classical-quantum channels,'' in \emph{Proc. of IEEE
  International Symposium on Information Theory}, Barcelona, Spain, Jul. 2016,
  pp. 2064--2068.

\bibitem{Bullock2025Fundamental}
M.~S. Bullock, A.~Sheikholeslami, M.~Tahmasbi, R.~C. Macdonald, S.~Guha, and
  B.~A. Bash, ``Fundamental limits of covert communication over
  classical-quantum channels,'' \emph{IEEE Transactions on Information Theory},
  vol.~71, no.~4, pp. 2741--2762, Apr. 2025.

\bibitem{Bullock2020}
M.~S. Bullock, C.~N. Gagatsos, S.~Guha, and B.~A. Bash, ``Fundamental limits of
  quantum-secure covert communication over bosonic channels,'' \emph{IEEE
  Journal of Selected Areas in Communications}, vol.~38, no.~3, pp. 471--482,
  Mar. 2020.

\bibitem{Gagatsos2020Covert}
C.~N. Gagatsos, M.~S. Bullock, and B.~A. Bash, ``Covert capacity of bosonic
  channels,'' \emph{IEEE Journal on Selected Areas in Information Theory},
  vol.~1, no.~2, pp. 555--567, Aug. 2020.

\bibitem{Wang2022Towards}
S.-Y. Wang, T.~Erdo\u{g}an, and M.~R. Bloch, ``Towards a characterization of
  the covert capacity of bosonic channels under trace distance,'' in
  \emph{Proc. of IEEE International Symposium on Information Theory}, Helsinki,
  Finland, Jun. 2022, pp. 354--359.

\bibitem{Wang2024}
S.-Y. Wang, S.-J. Su, and M.~R. Bloch, ``Resource-efficient
  entanglement-assisted covert communications over bosonic channels,'' in
  \emph{Proc. of IEEE International Symposium on Information Theory}, Athens,
  Greece, Jul. 2024, pp. 3106--3111.

\bibitem{Arumugam2018a}
K.~S.~K. Arumugam and M.~R. Bloch, ``Covert communication over a k-user
  multiple access channel,'' \emph{IEEE Transactions on Information Theory},
  vol.~65, no.~11, pp. 7020--7044, Nov. 2019.

\bibitem{Cho2022Covert}
K.-H. Cho and S.-H. Lee, ``Covert communication over gaussian multiple-access
  channels with feedback,'' \emph{{IEEE} Wireless Communications Letters},
  vol.~11, no.~9, pp. 1985--1989, Sep. 2022.

\bibitem{Bounhar2023Mixing}
A.~Bounhar, M.~Sarkiss, and M.~Wigger, ``Mixing a covert and a non-covert
  user,'' in \emph{Proc. of IEEE International Symposium on Information
  Theory}, Taipei, Taiwan, Jun. 2023, pp. 2577--2582.

\bibitem{bounhar2024}
\BIBentryALTinterwordspacing
------, ``Whispering secrets in a crowd: Leveraging non-covert users for covert
  communications,'' 2024. [Online]. Available:
  \url{https://arxiv.org/abs/2408.12962}
\BIBentrySTDinterwordspacing

\bibitem{Cho2021Treating}
K.-H. Cho and S.-H. Lee, ``Treating interference as noise is optimal for covert
  communication over interference channels,'' \emph{{IEEE} Transactions on
  Information Forensics and Security}, vol.~16, pp. 322--332, Jul. 2021.

\bibitem{Arumugam2018b}
K.~S.~K. Arumugam and M.~R. Bloch, ``Embedding covert information in broadcast
  communications,'' \emph{IEEE Transactions on Information Forensics and
  Security}, vol.~14, no.~10, pp. 2787--2801, Oct. 2019.

\bibitem{Kibloff2019}
D.~Kibloff, S.~M. Perlaza, and L.~Wang, ``Embedding covert information on a
  given broadcast code,'' in \emph{Proc. of IEEE International Symposium on
  Information Theory}, Paris, France, Jul. 2019, pp. 2169--2173.

\bibitem{Tan2019}
V.~Y.~F. Tan and S.-H. Lee, ``Time-division is optimal for covert communication
  over some broadcast channels,'' \emph{IEEE Transactions on Information
  Forensics and Security}, vol.~14, no.~5, pp. 1377--1389, May 2019.

\bibitem{Lee2018a}
S.~H. Lee, L.~Wang, A.~Khisti, and G.~W. Wornell, ``Covert communication with
  channel-state information at the transmitter,'' \emph{IEEE Transactions on
  Information Forensics and Security}, vol.~13, no.~9, pp. 2310--2319, Sep.
  2018.

\bibitem{ZivariFard2020}
H.~Zivari-Fard, M.~Bloch, and A.~Nosratinia, ``Keyless covert communication via
  channel state information,'' \emph{IEEE Transactions on Information Theory},
  vol.~68, no.~8, Aug. 2022.

\bibitem{Sobers2017}
T.~V. Sobers, B.~A. Bash, S.~Guha, D.~Towsley, and D.~Goeckel, ``Covert
  communication in the presence of an uninformed jammer,'' \emph{IEEE
  Transactions on Wireless Communications}, vol.~16, no.~9, pp. 6193--6206,
  Jun. 2017.

\bibitem{Hou_Kramer2014}
J.~Hou and G.~Kramer, ``Effective secrecy: Reliability, confusion and
  stealth,'' in \emph{2014 IEEE International Symposium on Information Theory},
  2014, pp. 601--605.

\bibitem{Hou2014}
------, ``Effective secrecy: Reliability, confusion and stealth,'' in
  \emph{Proc. of IEEE International Symposium on Information Theory}, Honolulu,
  HI, July 2014, pp. 601--605.

\bibitem{Tahmasbi2017c}
M.~Tahmasbi and M.~R. Bloch, ``Covert secret key generation,'' in \emph{Proc.
  of IEEE Conference on Communications and Network Security, Workshop on
  Physical-Layer Methods for Wireless Security}, Las Vegas, NV, Oct. 2017, pp.
  540--544.

\bibitem{Tahmasbi2018b}
------, ``Framework for covert and secret key expansion over classical-quantum
  channels,'' \emph{Physical Review A}, vol.~99, p. 052329, May 2019.

\bibitem{Tahmasbi2019a}
------, ``Covert and secret key expansion over quantum channels under
  collective attacks,'' \emph{IEEE Transactions on Information Theory},
  vol.~66, no.~11, pp. 7113--7131, Nov. 2020.

\bibitem{Lin2020Stealthy}
P.-H. Lin, C.~R. Janda, E.~A. Jorswieck, and R.~F. Schaefer, ``Stealthy secret
  key generation,'' \emph{Entropy}, vol.~22, no.~6, p. 679, Jun. 2020.

\bibitem{shannon1961two}
C.~E. Shannon, ``Two-way communication channels,'' in \emph{Proceedings of the
  Fourth Berkeley Symposium on Mathematical Statistics and Probability, Volume
  1: Contributions to the Theory of Statistics}, vol.~4.\hskip 1em plus 0.5em
  minus 0.4em\relax University of California Press, 1961, pp. 611--645.

\bibitem{Weng2019Capacity}
J.-J. Weng, L.~Song, F.~Alajaji, and T.~Linder, ``Capacity of two-way channels
  with symmetry properties,'' \emph{{IEEE} Transactions on Information Theory},
  vol.~65, no.~10, pp. 6290--6313, Oct. 2019.

\bibitem{Tekin2008General}
E.~Tekin and A.~Yener, ``The general {G}aussian multiple-access and two-way
  wiretap channels: Achievable rates and cooperative jamming,'' \emph{IEEE
  Transactions on Information Theory}, vol.~54, no.~6, pp. 2735--2751, Jun.
  2008.

\bibitem{Tekin2008erratum}
------, ``Correction to: ``the {G}aussian multiple access wire-tap channel''
  and ``the general {G}aussian multiple access and two-way wire-tap channels:
  Achievable rates and cooperative jamming'','' \emph{IEEE Transactions on
  Information Theory}, vol.~56, no.~9, pp. 4762--4763, Sep. 2010.

\bibitem{Pierrot2011a}
A.~J. Pierrot and M.~R. Bloch, ``{S}trongly secure communications over the
  two-way wiretap channel,'' \emph{IEEE Transactions on Information Forensics
  and Security}, vol.~6, no.~3, pp. 595--605, Sep. 2011.

\bibitem{ElGamal2013}
A.~{El Gamal}, O.~O. Koyluoglu, M.~Youssef, and H.~E. Gamal, ``Achievable
  secrecy rate regions for the two-way wiretap channel,'' \emph{IEEE
  Transactions on Information Theory}, vol.~59, no.~12, pp. 8099--8114, Dec.
  2013.

\bibitem{He2013Role}
X.~He and A.~Yener, ``The role of feedback in two-way secure communications,''
  \emph{IEEE Transactions on Information Theory}, vol.~59, no.~12, pp.
  8115--8130, Dec. 2013.

\bibitem{Helhal2018a}
N.~Helal, M.~Bloch, and A.~Nosratinia, ``Cooperative resolvability and secrecy
  in the cribbing multiple-access channel,'' \emph{IEEE Transactions on
  Information Theory}, vol.~66, no.~9, pp. 5429--5447, Sep. 2020.

\bibitem{ElGamal_Kim_2011}
A.~El~Gamal and Y.-H. Kim, \emph{Network Information Theory}.\hskip 1em plus
  0.5em minus 0.4em\relax Cambridge University Press, 2011.

\bibitem{gelfand2012calculus}
I.~Gelfand and S.~Fomin, \emph{Calculus of Variations}, ser. Dover Books on
  Mathematics.\hskip 1em plus 0.5em minus 0.4em\relax Dover Publications, 2012.

\bibitem{Hayashi2006}
M.~Hayashi, ``General nonasymptotic and asymptotic formulas in channel
  resolvability and identification capacity and their application to the
  wiretap channels,'' \emph{IEEE Transactions on Information Theory}, vol.~52,
  no.~4, pp. 1562--1575, April 2006.

\bibitem{Bloch2011e}
M.~R. Bloch and J.~N. Laneman, ``Strong secrecy from channel resolvability,''
  \emph{IEEE Transactions on Information Theory}, vol.~59, no.~12, pp.
  8077--8098, Dec. 2013.

\end{thebibliography}
